\documentclass[10pt,journal,compsoc]{IEEEtran}
\usepackage[ruled,vlined]{algorithm2e}
\usepackage{multirow}
\usepackage{textgreek}

\usepackage{siunitx}
\usepackage{hyperref}

 \usepackage{amsmath}
\ifCLASSOPTIONcompsoc
  \usepackage[nocompress]{cite}
\else
  \usepackage{cite}
\fi
\ifCLASSINFOpdf
   \usepackage[pdftex]{graphicx}
\else
\fi
\usepackage{array}
\usepackage{svg}
\usepackage{adjustbox}

\usepackage{booktabs,tabularx,ragged2e}

\newcolumntype{Y}{>{\Centering\hspace{0pt}}X}

\ifCLASSOPTIONcompsoc
  \usepackage[caption=false,font=footnotesize,labelfont=sf,textfont=sf]{subfig}
\else
  \usepackage[caption=false,font=footnotesize]{subfig}
\fi
\begin{document}
%
\title{An eXtended Reality Offloading IP Traffic Dataset and Models}
%
%
%
%

\author{Diego~Gonz\'alez~Mor\'in,~\IEEEmembership{Student Member,~IEEE,}
        Daniele~Medda,~\IEEEmembership{Student Member,~IEEE,}
        Athanasios~Iossifides,~\IEEEmembership{Member,~IEEE,}
        Periklis~Chatzimisios,~\IEEEmembership{Senior Member,~IEEE,}
        Ana~Garc\'ia~Armada,~\IEEEmembership{Senior Member,~IEEE,}        
        Alvaro Villegas,
        and Pablo~Per\'ez 
\IEEEcompsocitemizethanks{\IEEEcompsocthanksitem D.~Gonz\'alez~Mor\'in, P.~Per\'ez and Alvaro Villegas are with Nokia XR Lab, Nokia, Madrid - SPAIN \protect\\
E-mail: \{\url{diego.gonzalez_morin}, pablo.perez, alvaro.villegas\}@nokia.com
\IEEEcompsocthanksitem D. Medda and A. Iossifides are with the Department of Information and Electronic Engineering, International Hellenic University, Thessaloniki - GREECE. \protect \\
E-mail: \{dmedda, aiosifidis\}@ihu.gr
\IEEEcompsocthanksitem P. Chatzimisios is with the Department of Information and Electronic Engineering, International Hellenic University, Thessaloniki - GREECE and with the Department of Electrical and Computer Engineering, University of New Mexico, Albuquerque - USA. \protect \\
E-mail: pchatzimisios@ihu.gr
\IEEEcompsocthanksitem A.~Garc\'ia~Armada is with the Department of Signal Theory and Communications, University Carlos III of Madrid - SPAIN.  \protect \\
E-mail: agarcia@tsc.uc3m.es
}
}

%
%

\markboth{SUBMITTED TO IEEE TRANSACTIONS ON MOBILE COMPUTING, JANUARY 2023}%
{Shell \MakeLowercase{\textit{et al.}}: Bare Demo of IEEEtran.cls for Computer Society Journals}
%



\IEEEtitleabstractindextext{%
\begin{abstract}


In recent years, advances in immersive multimedia technologies, such as extended reality (XR) technologies, have led to more realistic and user-friendly devices. However, these devices are often bulky and uncomfortable, still requiring tether connectivity for demanding applications. The deployment of the fifth generation of telecommunications technologies (5G) has set the basis for XR offloading solutions with the goal of enabling lighter and fully wearable XR devices. In this paper, we present a traffic dataset for two demanding XR offloading scenarios that are complementary to those available in the current state of the art, captured using a fully developed end-to-end XR offloading solution. We also propose a set of accurate traffic models for the proposed scenarios based on the captured data, accompanied by a simple and consistent method to generate synthetic data from the fitted models. Finally, using an open-source 5G radio access network (RAN) emulator, we validate the models both at the application and resource allocation layers. Overall, this work aims to provide a valuable contribution to the field with data and tools for designing, testing, improving, and extending XR offloading solutions in academia and industry.
\\

\begin{center}
  \textcolor{red}{This work has been submitted to the IEEE for possible publication. Copyright may be transferred without notice, after which this version may no longer be accessible}  
\end{center}
\end{abstract}

\begin{IEEEkeywords}
EXtended Reality, 5G Networks, Offloading, Dataset, Traffic Models
\end{IEEEkeywords}}

\maketitle

\IEEEdisplaynontitleabstractindextext

%
\IEEEpeerreviewmaketitle

\IEEEraisesectionheading{\section{Introduction}\label{sec:introduction}}

%
%
%
%

\IEEEPARstart{u}{ndeniably}, the advances in immersive multimedia technologies introduced in the last five years are impressive. Extended reality (XR) technologies, which include virtual (VR) and augmented reality (AR) technologies, made huge leaps forward both in terms of realism and user interaction~\cite{suh2018state}. A significant factor in the current turmoil on the topic has undoubtedly been the remarkable interest of related companies such as Meta (formerly Facebook), Microsoft, and Sony. In particular, Meta has decided to focus heavily on the \textsl{metaverse} concept, thus, boosting the interest in these technologies, their potential use cases, and related issues~\cite{cheng2022will}. Due to their inherent characteristics, immersive use cases nowadays play a significant role in the development of a great multitude of enabling technologies. The recent interest in XR technologies has led to enormous investment increments, which have enabled lighter and cheaper devices to reach unprecedented levels of resolution. For example, the Varjo XR-3~\footnote{https://varjo.com/products/xr-3/} head mounted display (HMD) provides a visual resolution of 70 pixels per degree, matching the human eye's resolution. 

Advanced XR requires not only ultra-realistic resolution but also the implementation or improvement of other algorithms that aim to enhance user experience~\cite{kilteni2012embodiment}. This goal requires complex and computationally expensive algorithms, such as semantic segmentation~\cite{hengshuang2018semantic, dai20173drecon}, to run in real time. Therefore, to expand the current limits of XR technologies, HMDs must have access to high-end hardware with powerful graphical processing units for ultra-realistic rendering supported by machine learning (ML) processes. For this reason, advanced XR HMDs such as the Varjo XR-3 are tethered, uncomfortable and expensive. Consequently, next-generation wireless systems, such as 5G and 6G, must support XR technologies that have, as a whole, quickly become one of the killer use case families~\cite{perez2022emerging}. The goal, toward this end, is to offload XR heavy processing tasks to a nearby server, or a multi-access edge computing (MEC) platform, to loosen the in-built hardware requirements of XR HMDs while increasing their overall computing capabilities. However, XR offloading is a complex task with extreme requirements in terms of latency and throughput~\cite{gonzalez2022toward,gonzalez2020cutting}, which requires a well-designed and configured network. 

It is not trivial for developers and researchers to have access to fully developed XR offloading implementations. The current trend is to rely on pre-recorded or modeled traffic data, which are then fed to various simulation environments or actual wireless access network deployments. Pre-recorded traffic traces allow using extremely realistic data with simple use case-agnostic tools, such as tcpreplay~\footnote{https://tcpreplay.appneta.com/}. On the other hand, traffic models allow the generation of longer traffic traces while providing greater flexibility than pre-recorded traffic data. Even though it is true that the traffic characteristics for each XR use case can be very diverse, thus making it difficult to define a general-purpose model, access to modeled or pre-recorded XR traffic data can considerably accelerate and simplify the testing and prototyping steps. 




A number of previous works deal with immersive multimedia traffic capture and modeling or present ready-to-use models. Authors in~\cite{navarro2020survey} provide details on specific use cases employing AR and VR and how one can approximately model their behaviors using the models from 5G-PPP~\cite{osseiran2014scenarios,fantastic5g}. In~\cite{schulz2021analysis}, the authors modeled augmented reality downlink traffic using a classical two-state Markovian process. In~\cite{lecci2021open}, a complete framework aimed to model XR application is presented, alongside an accurate statistical analysis and an ad-hoc traffic generator algorithm. Furthermore, the work carried out in~\cite{lecci2021open} has been exploited to create a VR traffic generator framework for the ns-3 simulator~\cite{lecci2021ns3}. Authors in~\cite{bojovic2022enabling} model 3GPP-compliant traffic cases for next-generation mobile network applications, which include advanced gaming, but no explicit XR case is considered. Generally speaking, the previous works mostly focus on providing models for the downlink traffic. However, as described in~\cite{gonzalez2022toward,gonzalez2020cutting}, advanced XR technologies require multiple complex algorithms to run simultaneously in order to provide the user with a sufficiently high level of interaction, immersiveness, and experience. These algorithms, in many cases, require high-end hardware and therefore, can be considered potential offloading candidates. They require to be continuously fed with the sensor streams captured by the XR HMD, which can be as heavy as or more than ultra-realistic rendered frames. Aligned with this idea, 3GPP has recently included very detailed traffic models for AR and XR in Release 17, differentiated according to the type of data streamed~\cite{3gpp_17}. While the considered VR use cases are still centered only on distributed rendering solutions with a special focus on downlink traffic, AR traffic models also consider complex and heavy uplink traffic. 



In this work, we aim to provide realistic traffic traces and their associated models for two separate state-of-the-art XR offloading scenarios, both for downlink and uplink. Our goal is to complement and improve the models proposed in~\cite{3gpp_17}. First, the proposed scenarios, complement the ones described in~\cite{3gpp_17,lecci2021open}. Besides this, we provide the raw data, uploaded to~\cite{opensourcefikore}, which can be useful for many researchers not only to use it as it is for simulation or prototyping purposes but for generating other models more suitable for their use. We also provide a set of XR traffic models obtained from the traces. Differently from other models in the state of the art, we also model the inter-packet arrival time, which, as we show, can be extremely relevant for XR offloading resource allocation algorithms design. 

The scenarios under consideration include full XR offloading and egocentric human segmentation, both sitting on the very edge of the current state of the art. Therefore, we believe that our contribution will provide valuable tools to design, test, improve or extend wireless network solutions both in academia and industry. To our knowledge, this is the first work that provides both an accurate traffic dataset and validated models for the mentioned use cases and applications. Our main contributions can be summarized as: 
\begin{itemize}
    \item An XR offloading traffic dataset for two different relevant offloading scenarios, captured for multiple streaming resolutions; 
    \item XR traffic models obtained from the captured traces, including the inter-packet arrival time, not available in most of the models provided in the state of the art;
    \item A thorough validation of the proposed models using a realistic 5G radio access network (RAN) emulator, showing how an accurate inter-packet arrival time can considerably improve the quality of the models for specific applications. 
\end{itemize}

The remaining of this paper is organized as follows: Section~\ref{sec:scenarios} summarizes the two reference XR offloading scenarios; Sections ~\ref{section:arch} and ~\ref{section:capture} describe the offloading architecture and the traffic capture methodology employed in the use cases, respectively; in Section~\ref{section:modelling} we focus on the statistical modeling of the cases by using the previously captured traffic; furthermore, in Section~\ref{section:traffic_generation} we summarize artificial traffic generation with the use of the developed statistical models that we employ, in Section~\ref{section:validation}, in validation experiments that are carried out by means of simulation, in order to verify the behavioral compliance of the modeled traffic with the real captured one; lastly, final conclusions are drawn in Section~\ref{section:conclusion}.

\section{XR Offloading Scenarios}
\label{sec:scenarios}

Our goal is to capture a relevant IP traffic dataset for two demanding XR offloading scenarios, that is, full XR offloading (scenario A) and egocentric human segmentation algorithm offloading (scenario B). In scenario A, all the processing but the sensor capture is moved from the XR device to a nearby server. Differently from~\cite{3gpp_17}, we consider the VR HMD to be a very light device in charge of only capturing the sensor data. The sensor data are streamed to the server, where they get processed. The sensor info is used to render a new high-definition VR frame which is sent back to the device. This is a very relevant use case for advanced and future networks, which can enable ultra-light and wearable XR devices. In our case, we consider the sensor data to be generated by a stereo camera feed and inertial sensors. The inertial sensors traffic can be neglected as its associated throughput is much lower than the stereo camera feed throughput~\cite{gonzalez2022toward,gonzalez2020cutting}. This is an extremely demanding use case as the round trip times should lay below the frame update period, i.e., around 11 ms for a device running at 90~Hz. While there are some techniques to slightly expand this time budget, such as XR time warp~\cite{waveren2016timewarp}, the latency requirements are still tight, especially for ultra-high definition XR scenes rendering, encoding, and transmission. 

Scenario B focuses on the particular case of egocentric body segmentation~\cite{gonzalez2022segmentation}, since this is a promising state-of-the-art solution for XR applications. The upstream traffic includes the stereo camera traffic while the server is sending back simple binary masks to the device in which the white pixels correspond to the user's body. The received masks are used by the XR device to render only the pixels corresponding to the user's body within the VR scene. While still a demanding offloading use case, the overall requirements are much lower than in Scenario A, since the downlink stream is just composed of single-channel binary masks.

\section{Offloading Architecture}
\label{section:arch}

\begin{figure}[t]
\centering
\includegraphics[width=0.8\linewidth]{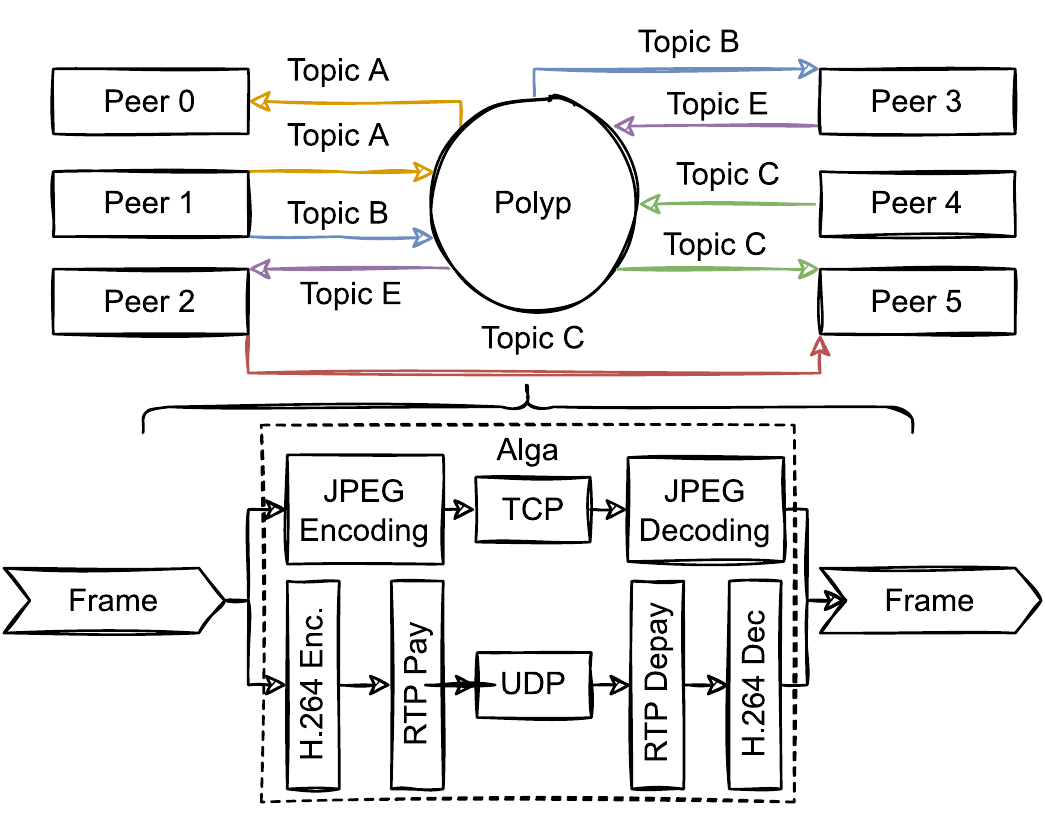}
\caption{The proposed offloading architecture strategy and simplified data flow for a general multi-peer scenario (top). Alga data flow for both TCP-JPEG and RTP-H-264 implemented transmission pipelines (bottom).  }
\label{fig:polypalga} 
\end{figure}

Our offloading architecture, described in~\cite{gonzalez2022arch}, relies on two main agents to share data between different peers. On one hand, we have Alga, which connects individual peers. On the other hand, we have Polyp, a data re-router, and replicator, in charge of transmitting the data from one source to one or multiple listening peers. We implemented a publisher-subscriber approach based on topics. When a client subscribes to a topic, Polyp is in charge of re-routing and replicating all the data of the topic toward this client. Similarly, when a client publishes data to a topic, Polyp ensures that these data are transmitted to all the peers subscribed to this topic. Our architecture allows direct communication between end clients without having to use Polyp. Polyp itself is a peer that can subscribe or publish to a topic. Alga is in charge of creating all the necessary connections and transmitting the data. The general representation of our architecture is depicted in Fig.~\ref{fig:polypalga}. 

The first version of this offloading architecture, implemented Alga using TCP for IP traffic transmission. Besides, to efficiently avoid TCP disadvantages, we sent each frame separately encoded in JPEG. This architecture served us to use our ML egocentric body segmentation algorithm, running on a nearby server, with a commercial XR HMD, the Meta Quest 2~\footnote{https://www.meta.com/en/quest/products/quest-2/}. However, joint JPEG encoding and TCP transmission, while useful in many scenarios due to their associated reliability, as described in~\cite{gonzalez2022arch}, were not originally designed to support high throughput and low latency. Therefore, we extended Alga's functionality incorporating H.264 video encoding~\cite{h264} and RTP (real-time transport protocol) over UDP transmission~\cite{rtp}. To encode the sensor streams in H.264 and pack the data in RTP frames the architecture uses GStreamer~\footnote{https://gstreamer.freedesktop.org/}. 

For traffic control reasons and to preserve compatibility with Polyp's in-built functionalities, we need to have control over the individual video frames and attach the metadata associated with them, such as the destination topic, timestamps, etc. This metadata can also be useful for performance analysis or bottleneck detection. To achieve this goal, we use RTP extended headers. Thus, the metadata is added to each video frame as an RTP extended header, which can be decoded and read on the receiving end. This is achieved using GStreamer in-built functionality. Alga's data flow for both TCP and RTP/UDP modes is depicted in Fig.~\ref{fig:polypalga}.

From the sending peer, the frames are fed to Alga in raw RGB format. Alga injects the raw frame along with its associated metadata into the GStreamer encoding and transmitting pipeline. If there are multiple peers subscribed to the same topic, the traffic is replicated and routed by Polyp, leaving this traffic untouched and just accessing the headers to read the target destination. In both this case and the case of direct traffic transmission between end peers using just Alga, the GStreamer pipeline receives and decodes the RTP frame. Once decoded, the frame can be accessed by the application layer. 

\section{Traffic Capture Methodology}
\label{section:capture}

As described in Section~\ref{section:arch}, our offloading architecture implementation has already been tested on a full end-to-end offloading solution using a commercial XR device, the Meta Quest 2. However, we decided to use a high-end laptop to emulate the XR offloading IP traffic for two main reasons: 
\begin{itemize}
    \item \textbf{Uncontrollable overhead} -- our architecture is optimized for wireless offloading via WiFi or advanced RAN networks such as 5G. We need to capture the data on the transmitting peer to avoid any overhead introduced by the wireless transmission, traffic routing, congestion, etc. These potential sources of overhead can lead to latencies, jitter, or packet loss which strongly depend on the used configuration, wireless technology, and other external factors. It is out of the scope of this work to model the network behavior and its associated configuration. However, we could not find an efficient manner to capture the IP traffic being transmitted from the XR device. 
    \item \textbf{Cover demanding XR offloading use cases} -- the Meta Quest 2 is not capable of handling demanding XR offloading use cases due to its limited computation capabilities. Our target is to cover XR offloading use cases which are still not possible with current XR or wireless access points technologies. 
\end{itemize} 
Following these considerations, all the data were captured using a high-end laptop, with an Intel Core i7-10870H CPU @ 2.20~GHz \texttimes\ 16, and 16 GB of RAM, running Ubuntu 18.04 LTS. The offloading architecture was set up and configured identically to an actual XR offloading deployment. For simplification and as we were not using an actual XR HMD, the IP traffic from each stream (scenario A and scenario B uplink and downlink streams) was captured via separate capture runs. Therefore, we used prerecorded data to capture the IP traffic. According to the scenarios described in Section~\ref{sec:scenarios} we recorded data for the following streams: 
\begin{figure}[th]
\centering
\includegraphics[width=\linewidth]{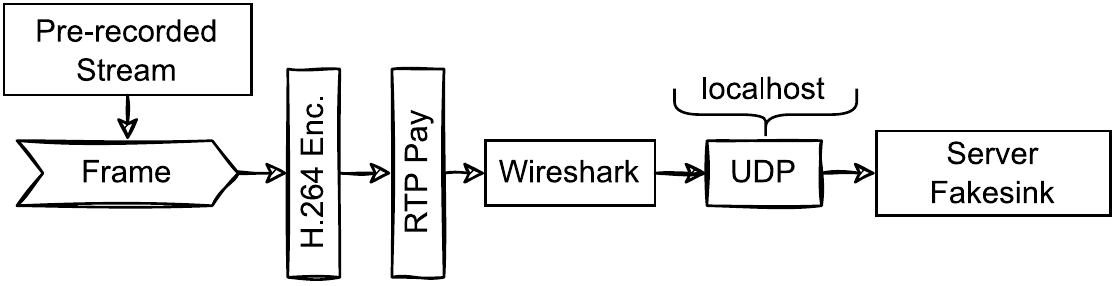}
\caption{ Simplified representation of the final setup used to capture the XR offloading IP traffic dataset.}
\label{fig:capturesetup} 
\end{figure}
\begin{itemize}
    \item \textbf{Stream 1 -- Uplink stereo camera stream}: This corresponds to the frontal stereo camera data, which are transmitted in both offloading scenarios A and B. The recorded data were obtained from the same stereo camera used in the end-to-end offloading solution of~\cite{gonzalez2022segmentation}. We recorded a continuous stereo video stream of 2560 \texttimes\ 960 resolution at 60~Hz, the maximum supported by the camera. While XR devices are expected to run at rates above 90~Hz, the sensor data are not required to be updated so fast~\cite{gonzalez2022toward,gonzalez2020cutting}. The prerecorded data had a length of 15 minutes.   
    \item \textbf{Stream 2 -- Downlink rendered frames}: This corresponds to the immersive frames rendered on the server in scenario A. In this case, we used a high-definition stereo video from a first person video game. The recorded video has a resolution of 3840 \texttimes\ 1920 and an update rate of 90~Hz. 
    \item \textbf{Stream 3 -- Downlink segmentation masks}: This corresponds to the binary pixel classification output by the egocentric body segmentation ML algorithm in scenario B. From the stream 1, we estimated the black and white binary single channel masks for each frame using the segmentation algorithm described in~\cite{gonzalez2022realtime}. Therefore, the resolution and update rate is the same as in stream 1 (2560 \texttimes\ 960 @ 60~Hz). 
\end{itemize}
To expand and add extra value to the presented dataset, we downscaled the three streams to different sets of resolutions/update rates that can be useful for potential researchers and applications. A summary of all the resolutions and update rates we used to generate the traffic data is shown in Table~\ref{tab:scenarios_resolutions}.  
\begin{table}[t]
    \centering
    \caption{Uplink and downlink resolutions and frame rates used to generate the proposed XR IP traffic dataset }
    \label{tab:scenarios_resolutions}
    \begin{tabular}{cccc}
         & \multicolumn{3}{c}{Resolution @ FPS}\\
         \midrule
         & Stream 1 & Stream 2 & Stream 3\\
         \midrule
    High & 2560 \texttimes\ 960 @ 60 & 3840 \texttimes\ 1920 @ 90 & 2560 \texttimes\ 960 @ 60\\
    Medium & 1920 \texttimes\ 720 @ 60 & 3840 \texttimes\ 1920 @ 72 & 1920 \texttimes\ 720 @ 60\\
    Low & 1280 \texttimes\ 480 @ 60 & & 1280 \texttimes\ 480 @ 60 \\    
    \bottomrule\\

    \end{tabular}
    
\end{table}
Each of the resolution/frames per second (FPS) and transmission direction (uplink or downlink) streams was captured separately. To capture the IP traffic, the client reads the individual raw frames, one by one, from the selected stream and sends them using the described architecture. As both the client and server run on the same machine to accelerate and simplify the capture process, we set a streaming client connected to a server, which just discards the incoming packets using GStreamer's \textit{Fakesink} module to avoid any additional overhead. There is no instance of Polyp and both client and server are directly connected using Alga in H.264-RTP mode: the raw frames are encoded using H.264 and packetized as RTP frames to be transmitted via UDP, in localhost. The IP traffic was captured using Wireshark, generating an individual packet capture (PCAP) file for each capture run. The final simplified capturing setup is depicted in Fig.~\ref{fig:capturesetup}. Each capture run had a duration of 10 minutes, for a total of 110 minutes of data.

\section{Traffic Modeling}
\label{section:modelling}
\begin{figure}[t]
\centering
\includegraphics[width=0.85\linewidth]{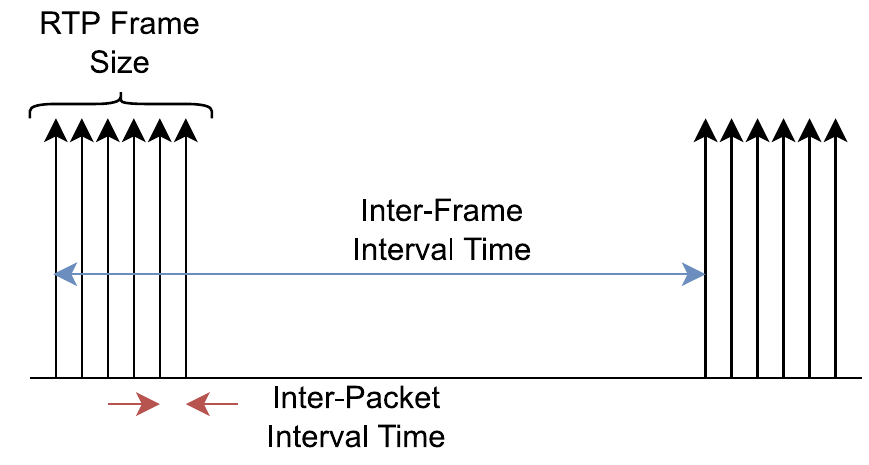}
\caption{Simplified IP packets (black arrows) representation packed in several RTP frames. The RTP frame size, inter-frame, and inter-packet interval times are illustrated.  }
\label{fig:rtpframesarrows}
\end{figure}

\begin{table}[th]
\centering
\caption{RTP frame size, inter-frame interval, inter-packet interval and individual IP packet sizes basic statistics from the captured data}
\label{tab:mean_values}
\begin{tabular}{l l c c c}
\multicolumn{5}{c}{Stream 1 -- Uplink stereo camera} \\
\midrule
                                        &Resolution      & Mean      & Std. Dev. & 95\textsuperscript{th} perc.  \\
\midrule
\multirow{3}{*}{\begin{tabular}[c]{@{}l@{}}Frame size \\ (bytes) \end{tabular}}     & Low  & 34602.44  & 9529.36  & 55735     \\
                                        & Medium  & 86149.87  & 19936.04 & 132384    \\
                                        & High & 232084.33 & 28141.99 & 269008    \\
\midrule
\multirow{3}{*}{\begin{tabular}[c]{@{}l@{}}Inter-frame \\interval (ms)\end{tabular}}   & Low  & 16.76     & 0.26     & 17.12     \\
                                        & Medium  & 16.76     & 0.50     & 17.53     \\
                                        & High   & 16.80     & 2.57     & 21.29     \\
\midrule
\multirow{3}{*}{\begin{tabular}[c]{@{}l@{}}Inter-packet \\interval (\textmu s)\end{tabular}} & Low  & 3.94      & 6.08     & 17.10     \\
                                        & Medium  & 3.53      & 5.47     & 17.27     \\
                                        & High   & 4.55      & 11.02    & 6.43      \\
\midrule
\multirow{3}{*}{\begin{tabular}[c]{@{}l@{}}IP packet size \\(bytes)\end{tabular}}         & Low  & 1280.79   & 356.58  & 1428      \\
                                        & Medium  & 1364.81  & 244.83 & 1428      \\
                                        & High   & 1403.88  & 154.31 & 1428    \\
\bottomrule
\\
\multicolumn{5}{c}{Stream 2 -- Downlink rendered frames}  \\
\midrule

                                        &Update rate      & Mean      & Std. Dev. & 95\textsuperscript{th} perc.  \\
\midrule
\multirow{2}{*}{\begin{tabular}[c]{@{}l@{}}Frame size \\ (bytes) \end{tabular}}     & 72~Hz  & 207968.42  & 122929.70  & 396402     \\
                                        & 90~Hz  & 163548.89  & 116837.86 & 339396    \\
\midrule
\multirow{2}{*}{\begin{tabular}[c]{@{}l@{}}Inter-frame \\interval (ms)\end{tabular}}   & 72~Hz  & 13.88      & 0.05     & 13.94     \\
                                        & 90~Hz  & 11.11      & 0.04      & 11.17     \\
\midrule
\multirow{2}{*}{\begin{tabular}[c]{@{}l@{}}Inter-packet \\interval (\textmu s)\end{tabular}} & 72~Hz  & 3.41      & 9.18     & 4.85      \\
                                        & 90~Hz  & 3.66     & 9.08     & 6.91     \\
\midrule
\multirow{2}{*}{\begin{tabular}[c]{@{}l@{}}IP packet size \\(bytes)\end{tabular}}         & 72~Hz  & 1400.04    & 171.38  & 1428      \\
                                        & 90~Hz  & 1392.66  & 191.91 & 1428      \\
\bottomrule
\\
\multicolumn{5}{c}{Stream 3 -- Segmentation masks} \\
\midrule
                                        &Resolution      & Mean      & Std. Dev. & 95\textsuperscript{th} perc.  \\
\midrule
\multirow{3}{*}{\begin{tabular}[c]{@{}l@{}}Frame size \\ (bytes) \end{tabular}}     & Low  & 4968.50  & 2175.03  & 7708     \\
                                        & Medium  & 8273.98  & 3921.00 & 13970    \\
                                        & High & 24378.90 & 11440.59 & 43458    \\
\midrule
\multirow{3}{*}{\begin{tabular}[c]{@{}l@{}}Inter-frame \\interval (ms)\end{tabular}}   & Low  & 16.76     & 0.20     & 17.05     \\
                                        & Medium  & 16.75     & 0.61     & 17.77     \\
                                        & High   & 17.10     & 3.30     & 22.44     \\
\midrule
\multirow{3}{*}{\begin{tabular}[c]{@{}l@{}}Inter-packet \\interval (\textmu s)\end{tabular}} & Low  & 7.01      & 6.17     & 15.04     \\
                                        & Medium  & 5.87      & 9.61     & 15.34     \\
                                        & High   & 7.54      & 24.80    & 24.63      \\
\midrule
\multirow{3}{*}{\begin{tabular}[c]{@{}l@{}}IP packet size \\(bytes)\end{tabular}}         & Low  & 749.83   & 517.39  & 1428      \\
                                        & Medium  & 933.53  & 527.48 & 1428      \\
                                        & High   & 1216.27  & 419.88 & 1428    \\
\bottomrule\\

\end{tabular}
\end{table}
In addition to releasing the PCAP files publicly, we made a systematic effort to statistically model the most relevant video streaming IP traffic parameters: i) RTP frame size, defined as the size of each individual RTP frame, ii) inter-frame interval, that is, the time between individual RTP frames and iii) inter-packet interval, i.e., the time between successive packets within an individual RTP frame. These parameters are depicted schematically in Fig.~\ref{fig:rtpframesarrows}. The main goal is to allow potential researchers, in the context of wireless communication systems analysis and evaluation, to generate realistic XR IP traffic, online or offline, based on the models derived.

\subsection{Data Pre-processing}
The PCAP files are large and contain a lot of information that can be useful in future works, such as the transmitted bytes themselves or other relevant metadata. To derive the traffic models, we store the payload, timestamps and the new RTP frame marker bit of the individual IP packets coming into the arbitrary port used for transmission. This bit information is necessary to identify a new frame. Data pre-processing takes place in two steps, as follows.

In the first step, we obtain a list of all the captured IP packets, ordered according to their timestamp. For each packet, we keep the payload in bytes, the timestamp, and a custom boolean indicator, i.e., a combination of the bit marker and the timestamp separation, which determines if the IP packet initiates a new RTP frame or not. This first pre-processing step is implemented using Python and Scapy\footnote{https://scapy.net/} library to parse the PCAP file. 

In the second step, we go through all the IP packets and group them in individual RTP frames according to the custom boolean indicator. Then we estimate, for each RTP frame, the total size in bytes (RTP frame size), the time in between consecutive frames (inter-frame intervals), and the time in between consecutive IP packets (inter-packet intervals). These parameters are stored in three separate arrays and saved as an NPY (Python NumPy format) file. These NPY files are the ones used to model the IP traffic. This second step is implemented in Python as well.

\begin{figure*}[t]
\centering
\includegraphics[width=\linewidth]{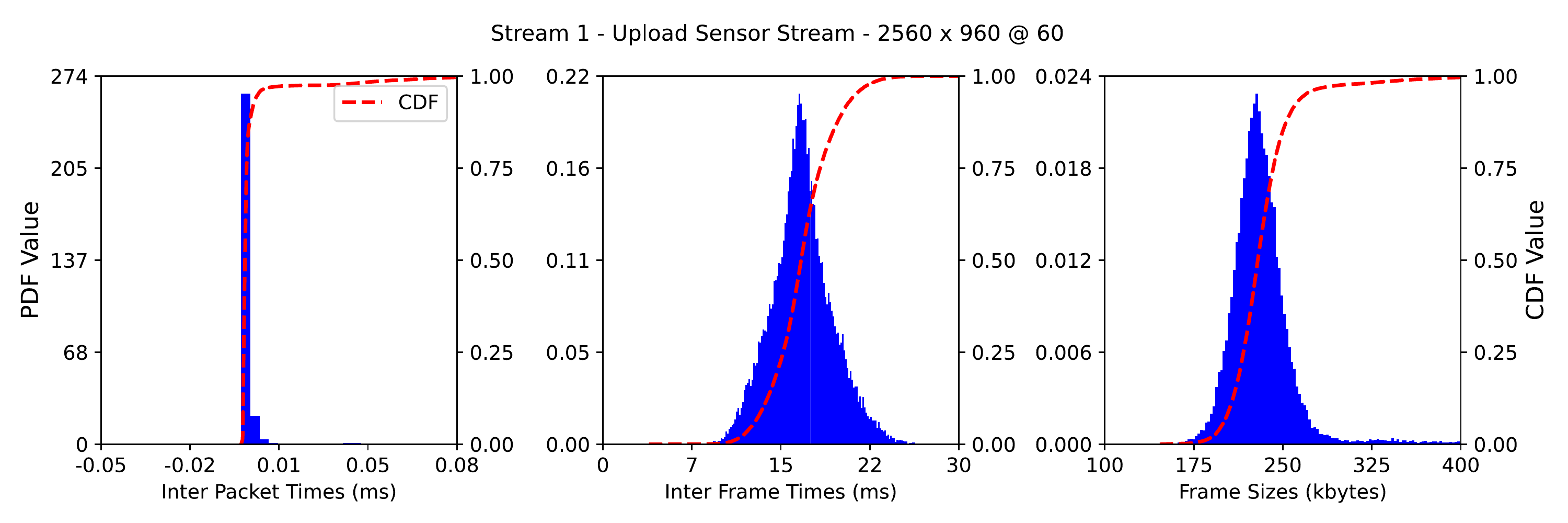} \\
\includegraphics[width=\linewidth]{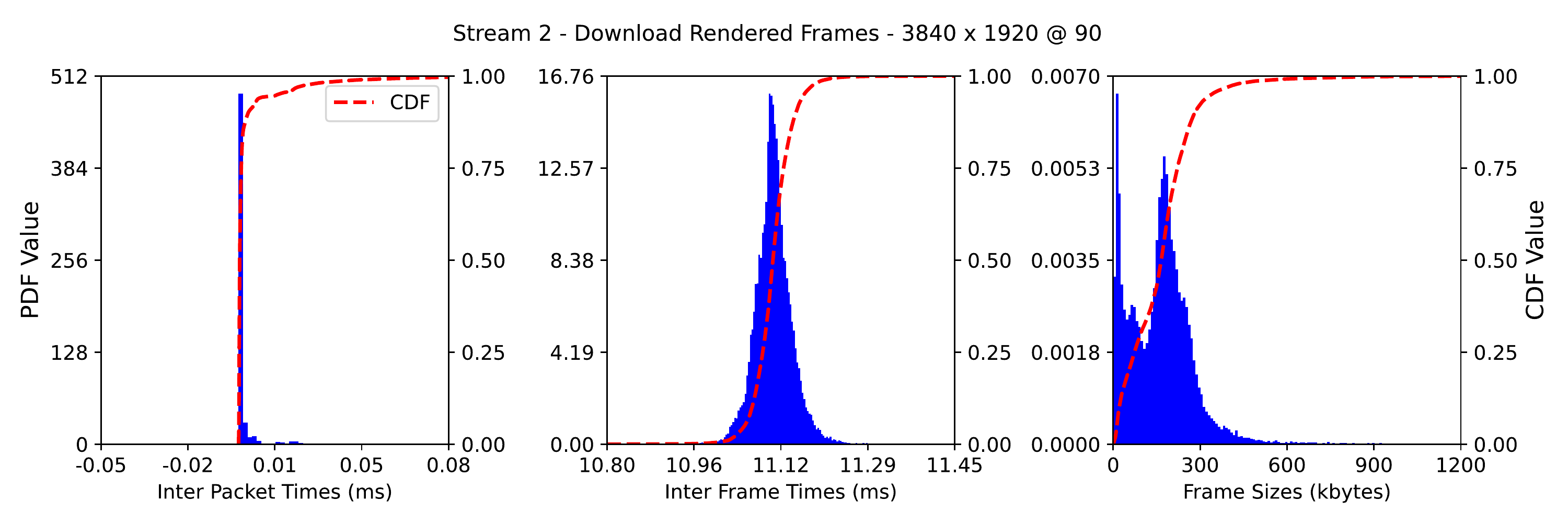}\\
\includegraphics[width=\linewidth]{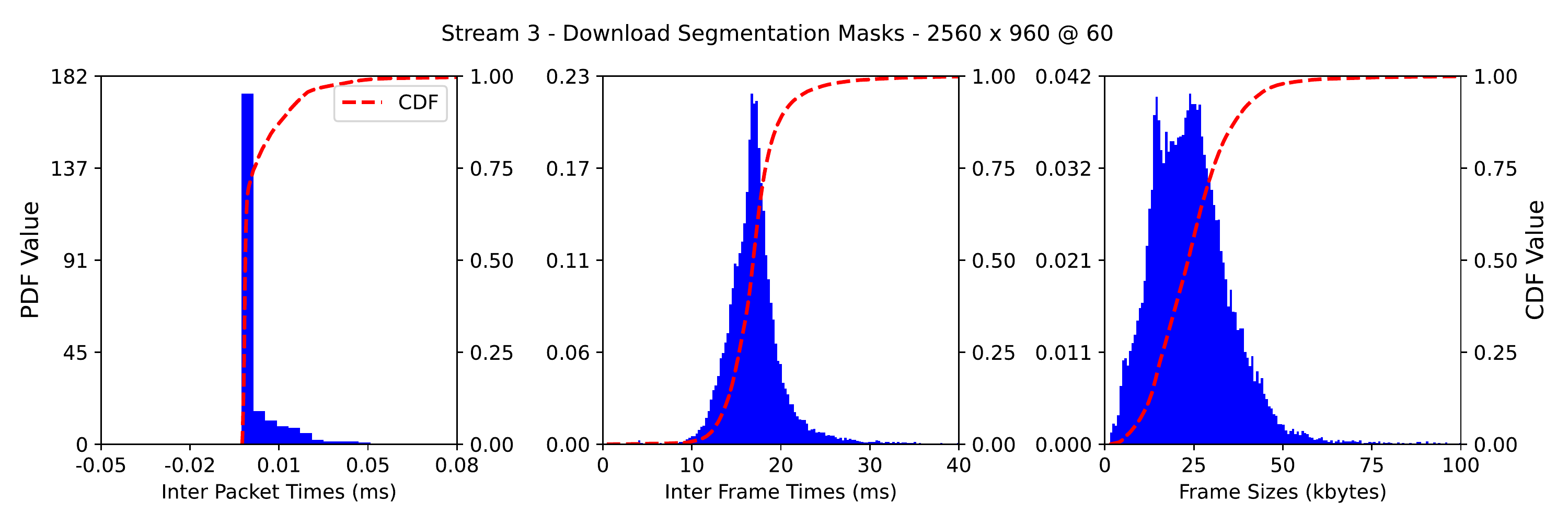}
\caption{Histograms (in blue) and cumulative distribution functions (CDF, in red) from the captured data for Stream 1 (top), 2 (middle), and 3 (bottom) for the target parameters: inter-packet interval times (left), inter-frame interval times (center) and frame sizes (right). }
\label{fig:example_hist} 
\end{figure*}

These two steps are applied to all the captured PCAP files. The final outputs are stored as individual files to easily identify each capture run. In Table~\ref{tab:mean_values} we show the basic statistics, i.e., mean value, standard deviation, and  95\textsuperscript{th} percentile of all the captured data cases, for the frame size, inter-frame interval, inter-packet interval, and IP packet size. The packet size information is useful to generate synthetic data from the fitted models. 

\subsection{Prior Data Analysis}
Before taking any modeling decisions, we studied the histograms of the pre-processed data. In particular, we plotted the histograms for all the parameters to be modeled for all the captured data. In Fig.~\ref{fig:example_hist} we present examples of the RTP frame size, inter-frame interval, and inter-packet interval histograms, for the high-resolution Stream 1 and 3 (at 60~Hz), as well as Stream~2 at 90~Hz. 

We observe, in all streams, that the inter-frame intervals are evenly distributed around a mean value that coincides with the frame update period according to the selected FPS value. Due to variable rate encoding, which guarantees low latency, the coding rate and the frame size may include peaks and variations. For the 60~Hz captured data, this is not an issue since the encoder is faster than the frame update period for all cases and resolutions. However, for very high resolutions and frame update rates, the coding rate needs to dynamically adapt, resulting in frame sizes with more than one peak, as shown for Stream 2 in Fig.~\ref{fig:example_hist}. This also affects the standard deviation of the inter-frame interval, which is reduced in Stream 2 cases due to the stricter encoding time requirements.

Regarding the potential distributions to model the target parameters, we observe, that in both Stream 1 and 3, these parameters can be modeled as unimodal continuous distributions. On the other hand, we observe that the distribution of the RTP frame sizes of Stream 2 presents two local maxima. These local maxima are smaller for the higher frame update rate (90~Hz) depicted in Fig.~\ref{fig:example_hist}. Nevertheless, we decided to model Stream 3 RTP frame sizes as continuous unimodal distributions as well and check if they provide a sufficiently good fitting before testing multimodal distributions. 

In Fig.~\ref{fig:example_hist} we can observe that the inter-packet interval distribution seems not to be unimodal, since slight changes in convexity appear. However, the inter-packet intervals lay in the order of the microsecond, as shown in Table~\ref{tab:mean_values}. On that scale, many external sources can affect the measured value, such as the operating system particular operations, Wireshark processing, etc. Again, modeling these possible external factors that can affect the inter-packet intervals is out of the scope of this work. Therefore, we choose to move forward with the simple approach of modeling the inter-packet interval time as a unimodal continuous distribution.  

\subsection{IP Traffic Models}

\begin{figure*}[ht]
\centering
\includegraphics[width=\linewidth]{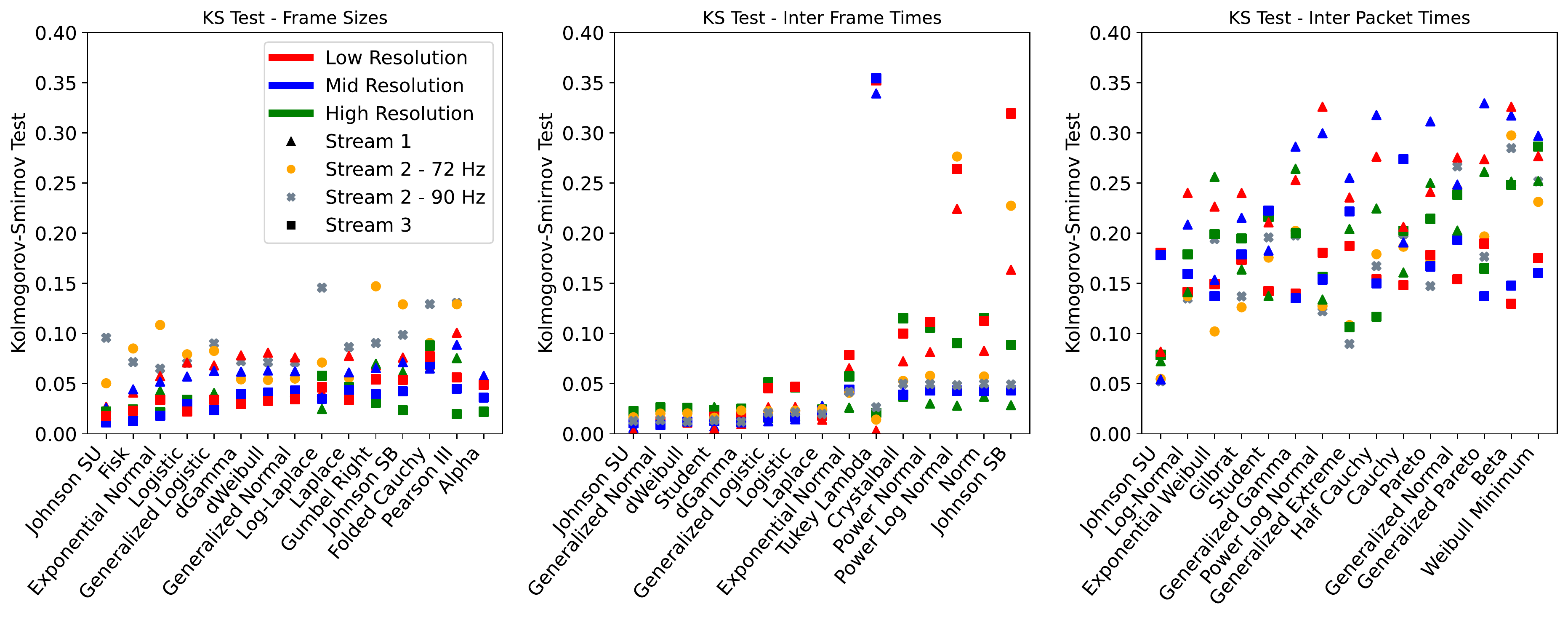}
\caption{Best KS test scoring models (from left to right) for the target parameters to be modelled: RTP frame sizes (left), inter-frame interval times (center), and inter-packet interval times (right).  }
\label{fig:ksvalues} 
\end{figure*}

\begin{figure}[t]
\centering
\includegraphics[width=\linewidth]{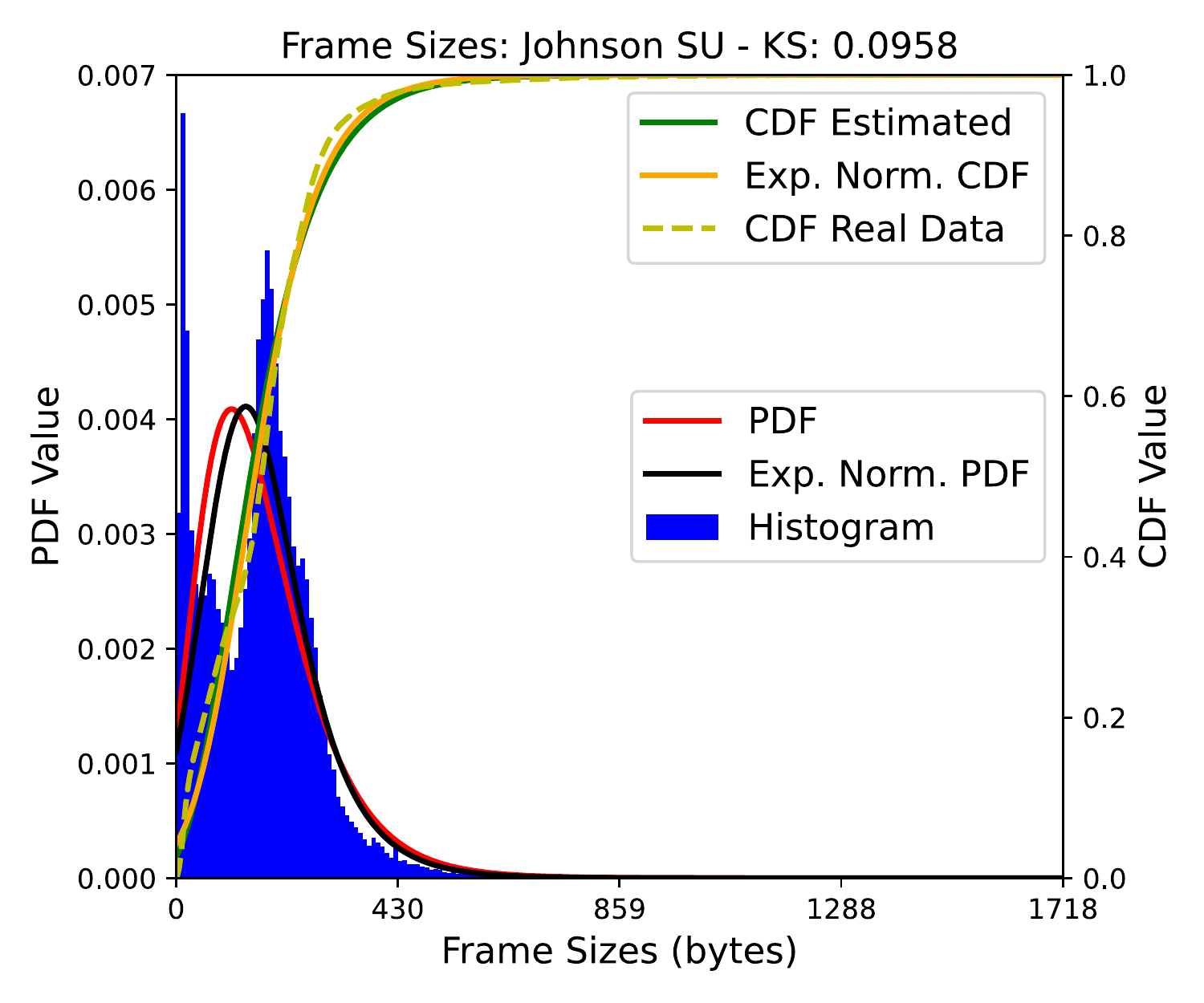}
\caption{Johnson's $S_U$ and exponential normal fitted distributions' probability density functions (PDF) and CDFs for the Stream 2 and 90~Hz case.}
\label{fig:normfsksvalues} 
\end{figure}

There is a wide range of well-established and commonly used continuous distributions for the parameters under consideration. To find the best candidate distributions that fit our data, we used Python's Scipy library\cite{virtanen2020scipy}. Scipy is capable of modeling more than 90 different continuous distributions. We decided to fit all the distributions available and evaluate their goodness of fit using the Kolmogorov-Smirnov (KS) test~\cite{kstest}. The KS test quantifies the distance between the empirical CDF $F_n(x)$ of a sample and the fitted CDF of an arbitrary distribution $F(x)$ as 

\begin{figure*}[t]
\centering
\includegraphics[width=\linewidth]{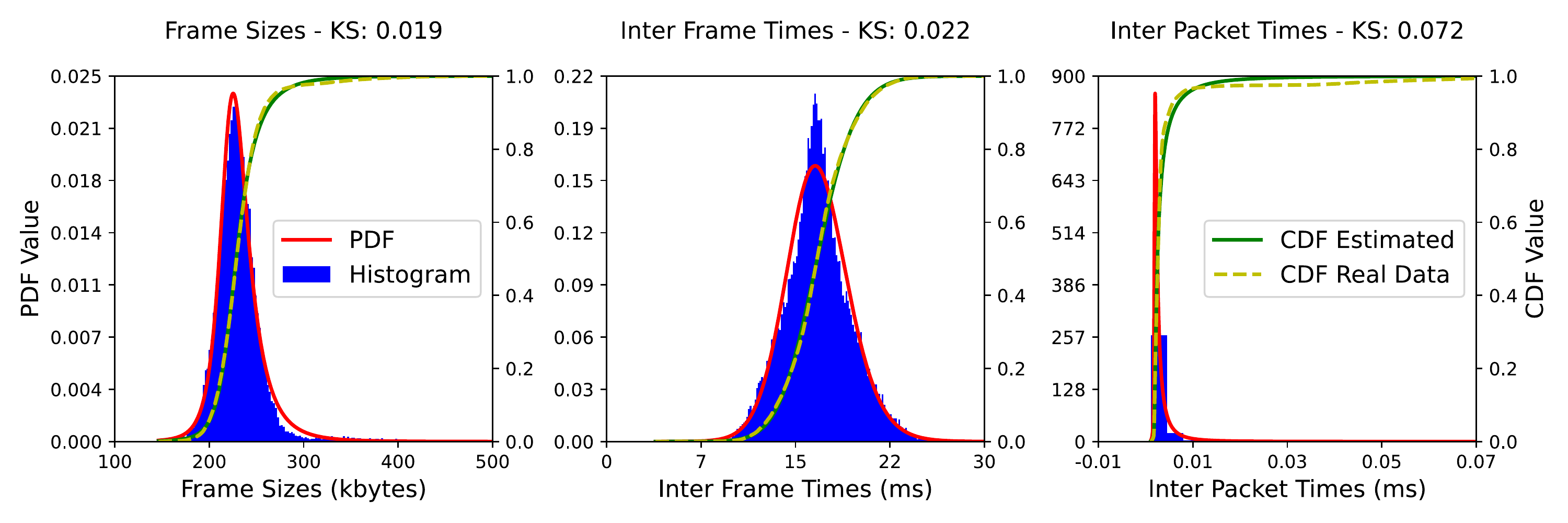}
\caption{Histograms (in blue) and CDFs (yellow) from the captured data for Stream 1 for the parameters: inter-packet interval times (left), inter-frame interval times (center) and frame sizes (right). On top, the Johnson's $S_U$ fitted distribution's PDF (red) and CDF (green).}
\label{fig:examplefit} 
\end{figure*}

\begin{equation}
    \text{KS} = \sup_{x}|F_n(x) - F(x)|, 
\end{equation}
where $\sup_{x}$ is the supremum of all the set of distances across $x$ values. The lower the KS test value, the better the fitting of the candidate distribution with the captured data. The KS test results of the 15 best-fitted distributions for each parameter and stream type are depicted in Fig.~\ref{fig:ksvalues}, sorted from best (left) to worst (right). We observe that Johnson's $S_U$ distribution~\cite{johnson1949johnsonsu} obtains the best mean KS value across all the captured data. This distribution was proposed by N.~L.~Johnson in 1949 and has been historically used in finances. The key characteristic of Johnson's $S_U$ distribution is its flexibility which originates from its four parameters that allow the distribution to be either symmetric or asymmetric. The probability density function (pdf) is expressed as 
\begin{equation}
\begin{split}
f(x, \gamma, \delta, \lambda, \varepsilon) &= \frac{\varepsilon}{\delta\cdot m(x,\gamma,\delta)} 
\phi \bigl\{\gamma + \delta\log(x) \\
& \quad \cdot\left [ k(x,\gamma,\delta) + m(x,\gamma,\delta) \right] \bigr\} ,
\end{split}
\end{equation}
where
\begin{equation}
k(x, \gamma, \delta) = \frac{x-\gamma}{\delta}, \quad m(x, \gamma, \delta) = \sqrt{k(x, \gamma, \delta)^2 + 1},
\end{equation}
with $\gamma$ and $\delta$ being the location and scale parameters, respectively, $\lambda$ and $\varepsilon$ the Johnson's $S_U$ specific shape parameters, and $\phi(\cdot)$ the pdf of the normal distribution. 

By further inspecting Fig.~\ref{fig:ksvalues}, we notice that for the RTP frame sizes, the only case that Johnson's $S_U$ does not provide the best fit is for the Stream 2 @ 90~Hz, for which the Exponential Normal distribution is the best. However, as we can see in Fig.~\ref{fig:normfsksvalues} the practical differences between the two distributions for Stream 2 @ 90~Hz are small enough. Besides, even if Johnson's $S_U$ fit is not as accurate as in the other RTP frame size distributions (see Fig.~\ref{fig:examplefit}), the measured KS values obtained are low enough, with a good fit for the larger packet sizes. Therefore, we decided to model and evaluate the RTP frame sizes using the Johnson's $S_U$ distribution for all the captured data. Similarly, we decided to use Johnson's $S_U$ distribution to model also the inter-frame and inter-packet intervals. The parameters of Johnson's $S_U$ distribution for all the traffic parameters under consideration and captured data are summarized in Table~A.1 of the Appendix.

\section{Realistic Traffic Generation}
\label{section:traffic_generation}

Our next goal is to build a tool that allows the generation of realistic XR offloading IP traffic. Such a tool is useful for researchers and application developers to generate and use synthetic data for analysis or incorporate it into complex link-level or system-level simulations. While other video XR traffic~\cite{3gpp_17} state-of-the-art models only consider the frame size and inter-frame interval for generating synthetic data, we believe that including the inter-packet interval data extends the applicability of our models to a wider range of research efforts. For instance, when designing novel or advanced resource allocation techniques, an accurate inter-packet interval model might be extremely useful and lead to better and more appropriate solutions. 

To create synthetic data we have to generate random values from the fitted distributions. Towards this end, we used Scipy's \textit{rvs} in-build function which generates random values from a specific distribution. In addition, we need the size of the individual RTP packets. In the real captured data this is not constant, as shown in Table~\ref{tab:mean_values}, in terms of the IP packet sizes, especially for Stream~3, since the way the segmentation mask is coded and organized in RTP packets varies from the regular color video stream (1 and 2). In general, the packets of each RTP frame have a fixed size chosen in the encoding/RTP framing pipeline (1442 bytes in our case). The first (including the RTP header) and the last are usually different. Depending on the chosen pipeline and configuration there may be smaller packets also in between, as in our case. However, these phenomena happen rarely as we can observe in the packet size histograms. The significant difference  between the mean and the maximum packet sizes in low throughput streams, such as Stream 3, is expected because the number of packets between the first and last within an RTP frame is small (smaller than 5 in the low resolution Stream 3 case). Therefore, we consider two IP packet size options: i) the mean size value, as in Table~\ref{tab:mean_values} or ii) the 95\textsuperscript{th} percentile value. We refer to case i) as \textit{Mean Packet} and case ii) as \textit{Max Packet}. 

\begin{algorithm}[t]
\caption{\label{alg:datagen}Synthetic IP packets generation algorithm.}
\SetKwInOut{Input}{input}\SetKwInOut{Output}{output}

\Input{$N_{RTP}$, $s_{RTP}$}
\Output{A sequence of IP packets}

\While{$N_{RTP}^{\text{generated}} < N_{RTP}$}
{
    1. $s_{IP} \gets \text{FS}_{\text{random}}$ \\
    2. $N_{IP} \gets s_{RTP} / s_{IP}$

    3. \While{$N_{IP}^\text{generated}  < N_{IP}$}
    {
        3.1 $\Delta t_{IP} \gets \text{IPI}_{\text{random}}$\\
        3.2 $t_s \gets t_s + \Delta t_{IP}$ \\
        3.3 Store new IP packet P($t_s$, $s_{RTP}$)\\
        3.4 $N_{IP}^{\textit{generated}} \gets N_{IP}^{\textit{generated}} + 1$
    }
    4. $\Delta t_{IF} \gets \text{IFI}_{\text{random}}$ \\
    5. $t_s \gets t_s + \Delta t_{IF}$ \\
    6. $N_{RTP}^{\textit{generated}} \gets N_{RTP}^{\textit{generated}} + 1 $
}
\end{algorithm}


Once we have the generators and packet sizes, we can easily define a procedure for synthetic realistic IP traffic generation, as described in Alg.~\ref{alg:datagen}. For each RTP frame among the $N_{RTP}$ to be generated, we begin by getting its size $s_{RTP}$ from the selected RTP generator, and by choosing the IP packet size $s_{IP}$ equal to Max Packet or Mean Packet. Then we compute the total number of packets $N_{IP}$ simply by dividing $s_{RTP}$ by $s_{IP}$. We continue by generating $N_{IP}$ packets of size $s_{IP}$, each with a specific timestamp $t_s$. The timestamp is computed by adding a random inter-packet interval $\Delta t_{IP}$ to the previous packet timestamp. Once all $N_{IP}$ packets are generated, a new randomly picked inter-frame interval $\Delta t_{IF}$ is added to the current timestamp. The random values are generated from the modelled distributions. The above procedure is repeated for each RTP frame.

The described algorithm can be easily implemented in any programming language and therefore used in any simulation environment. Additionally, it can be used to create synthetic traffic traces by storing the generated packets in a separate PCAP file and utilize them at a later time. 


\section{Validation Experiments}
\label{section:validation}
In this section, we test the traffic generated with the methodology described in the previous section, over a realistic RAN scenario, to determine its ability to accurately mimic the behavior of the captured XR offloading data traffic. To do this, we first compare the average throughput obtained from the captured data with the corresponding generated synthetic data obtained. Then, we study the behavior of the different synthetic data models in terms of application layer throughput and latency in the most relevant offloading scenarios using a real-time 5G RAN emulator. Finally, we thoroughly examine the impact of the type of traffic model used on  resource allocation by comparing synthetic data from different models with actual XR traffic.
\begin{table}[t]
\centering
\caption{Mean throughput of the generated synthetic data in comparison with the captured traffic's throughput }
\label{tab:generatedTp}
\begin{adjustbox}{width=\linewidth}
\begin{tabular}{lccccc}
      \multicolumn{6}{c}{Stream 1 -- Uplink stereo camera} \\
      \toprule
      & \multicolumn{1}{c}{Captured} & \multicolumn{2}{c}{Max Packet model} & \multicolumn{2}{c}{Mean Packet model}\\
      & (Mbps) & (Mbps) & Error (\%) & (Mbps) & Error (\%)\\
      \midrule
Low   & \multicolumn{1}{c}{16.51}  & 16.49     & \multicolumn{1}{c}{0.12}  & 16.49    & 0.12  \\
Med.   &\multicolumn{1}{c}{41.11}  &  41.15    & \multicolumn{1}{c}{0.09}  & 40.96    & 0.36  \\
High  & \multicolumn{1}{c}{110.55}  & 110.50   & \multicolumn{1}{c}{0.05}  & 110.27    & 0.25  \\
\bottomrule\\
      \multicolumn{6}{c}{Stream 2 -- Downlink rendered frames}    \\
      \toprule
      & \multicolumn{1}{c}{Captured} & \multicolumn{2}{c}{Max Packet model} & \multicolumn{2}{c}{Mean Packet model}\\
      & (Mbps) & (Mbps) & Error (\%) & (Mbps) & Error (\%)\\
      \midrule
72~Hz & \multicolumn{1}{c}{119.79}  & 121.36   & \multicolumn{1}{c}{1.29}  & 120.83   & 0.86  \\
90~Hz & \multicolumn{1}{c}{117.76}  & 117.74   & \multicolumn{1}{c}{0.02}  & 118.60     & 0.71  \\
\bottomrule\\
      \multicolumn{6}{c}{Stream 3 -- Downlink segmentation mask}  \\
      \toprule
      & \multicolumn{1}{c}{Captured} & \multicolumn{2}{c}{Max Packet model} & \multicolumn{2}{c}{Mean Packet model}\\
      & (Mbps) & (Mbps) & Error (\%) & (Mbps) & Error (\%)\\
      \midrule
Low   & \multicolumn{1}{c}{2.37}  & 2.35    & \multicolumn{1}{c}{0.89}  & 2.36     & 0.51  \\
Med.   &\multicolumn{1}{c}{3.95}  &  3.92     & \multicolumn{1}{c}{0.66}  & 3.94     & 0.35  \\
High  &\multicolumn{1}{c}{11.4}  &  11.44    & \multicolumn{1}{c}{0.38}  & 11.44    & 0.32 \\
\bottomrule
\end{tabular}
\end{adjustbox}
\end{table}

The first step is to compare the generated mean throughput of the synthetic data using the modeled Johnson's $S_U$ distribution with the captured data. The mean throughput results of the captured and synthetic data, for both Max Packet and Mean Packet cases (IP packet sizes), are shown in Table~\ref{tab:generatedTp}. The differences between the synthetic and captured data throughput are also included as percentage differences. We can observe that the throughput differences are low in all cases, with a peak of 1.29\% for Stream 2 and 72~Hz case. All other cases present differences below 1\%, for both IP packet sizes, with an average error of 0.435\%, and 0.438\%, for Max Packet, and Mean Packet case, respectively.

\begin{table}[t]
\centering
\caption{Common simulation parameters used in all the experiment runs}
\label{tab:simparameters}
\begin{tabular}{cc}
\multicolumn{2}{c}{Simulation Parameters} \\        
\toprule
TDD Configuration         & 1(UL):1(DL)           \\
Modulation                & 256-QAM               \\
Frequency Band            & 26.5 GHz              \\
UE MIMO Layers            & 2                     \\
Allocation Type           & 0                     \\
Allocation Configuration        & 1                     \\
Scenario                  & Rural Macrocell      \\
\bottomrule
\end{tabular}
\end{table}


The next evaluation step is to compare the behavior of both synthetic and captured data on a realistic advanced 5G RAN deployment. Towards this end, we used the open-source 5G RAN real-time emulator, named FikoRE~\cite{morin2022fikore,opensourcefikore}. FikoRE has been specifically designed for application layer researchers and developers to test their solutions on a realistic RAN setup. It supports the simulation of multiple background user equipment (UE) while handling high actual IP traffic throughput (above 1 Gbps). For our validation experiments, FikoRE runs as a simulator since we are not injecting actual IP traffic, but the traces from the captured or synthetic data. We tested the two scenarios described in Section~\ref{sec:scenarios}, with the following setup: 
\begin{itemize}
    \item \textbf{Scenario A -- Full Offloading}: On the downlink side, we chose to evaluate the 72~Hz rendered frames stream since it represents the current rendering offloading possibilities of commercial XR devices such as the Meta Quest 2. The Meta Quest~2 devices are capable of performing offloaded rendering, via a WLAN network, to a laptop in charge of rendering the immersive scene. The recommended setup is 72~Hz, being the rendering resolution of 1832 \texttimes\ 1920 per eye, which is slightly smaller than our captured data for the rendering frames stream. The uplink corresponds to the sensor stream (Stream~1) with a stereo resolution of 1920 \texttimes\ 720. 
    \item \textbf{Scenario B -- Egocentric Human Body Segmentation}: Successful deployment of this scenario was achieved in previous works~\cite{gonzalez2022segmentation}. While our deployment uses smaller resolutions, we evaluated the scenario in which both Stream~1 and 3 use a resolution of 1920 \texttimes\ 720.
\end{itemize}

\begin{figure}[t]
\centering
\includegraphics[trim={0 0 {1.94\linewidth} 0},width=\linewidth, clip]{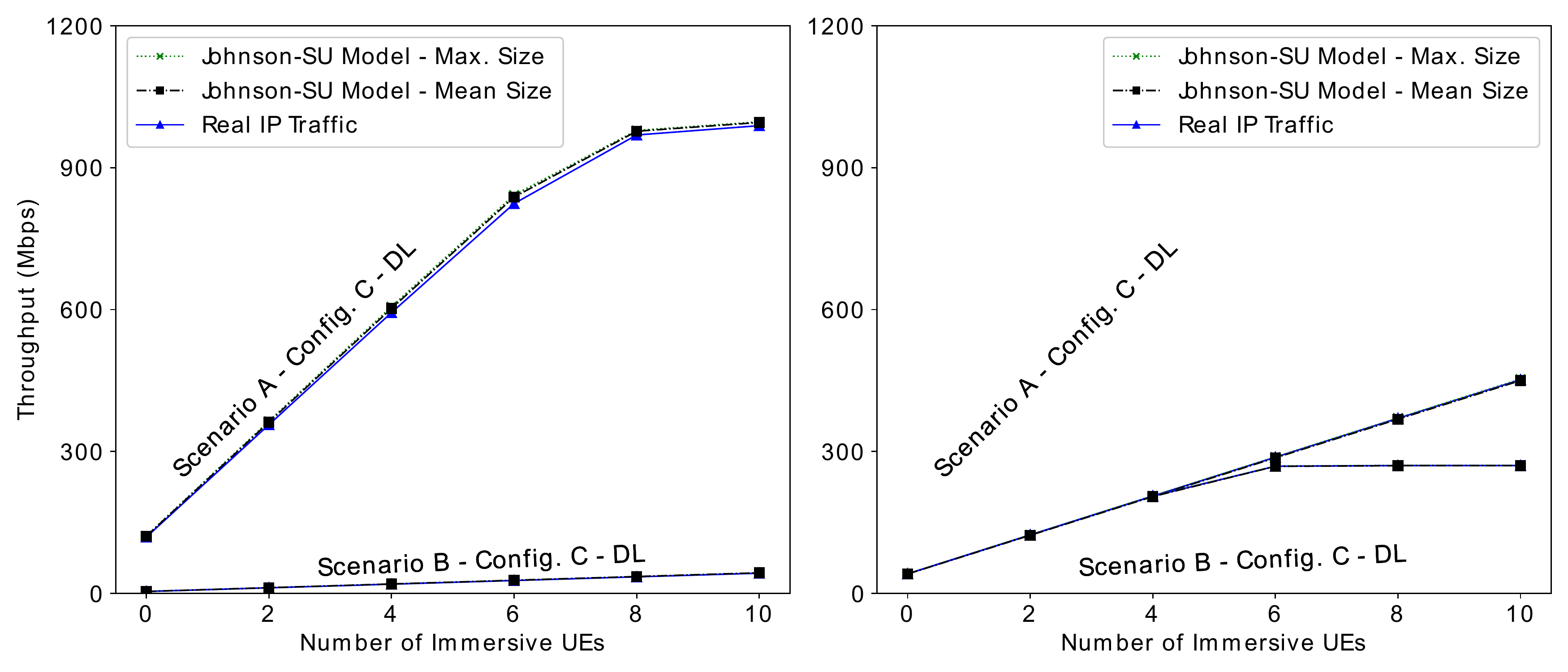}
\caption{Mean downlink throughput measured for Scenario A and B along with configuration C for the captured and synthetic data. }
\label{fig:resultstp} 
\end{figure}

Both offloading scenarios were evaluated in three different network configurations: 
\begin{itemize}
    \item \textbf{Configuration A -- Multiple background UEs and a single immersive UE with proportional fair (PF)}: In this scenario we simulated multiple UEs which are transmitting 5~Mbps of traffic in each direction. The throughput is not continuous, but is synthetically generated using the video streaming models from~\cite{tgaxevaluation} applicable for streaming applications such as Netflix. The emulator is set up with a single carrier of 100~MHz bandwidth on the 26.5~GHz millimeter wave (mm-wave) frequency band. Resource allocation takes place based on the PF metric~\cite{capozzi2013metrics}, using 1:1 (downlink:uplink) time division duplexing (TDD). We tested this network with a single immersive UE and 0, 20, 40, 60, 80 and 100 background UEs with 5~Mbps traffic in each direction. The network starts saturating around 80 simultaneous UEs. 
    \item \textbf{Configuration B -- Mutiple immersive UEs with PF}: In this scenario we have multiple immersive UEs, all using the same synthetic data. The throughput per UE is much higher than in Configuration~A, so we increased the total bandwidth to 200~MHz in order to be able to simulate more UEs before reaching network UE saturation. 
    \item \textbf{Configuration C -- Mutiple immersive UEs with maximum throughput (MT)}: this setup is identical to Configuration~B only changing the resource allocation metric used from PF to MT~\cite{capozzi2013metrics}. 
\end{itemize}
\begin{table*}[t]
\centering
\caption{Emulated application level throughput and latency comparison between captured and synthetic traffic. The synthetic experiments are repeated using the maximum and mean packet sizes}
\label{tab:macroresults}
\begin{tabular}{lcccccccc}
\multicolumn{9}{c}{Configuration A -- Multiple background UEs and a single immersive UE with PF}                                                                \\
              \cmidrule{2-9}
              & \multicolumn{4}{c}{Scenario A: Full offloading}                                    & \multicolumn{4}{c}{Scenario B: Deep learning offloading}                     \\
              \cmidrule{2-9}
              & \multicolumn{2}{c}{Throughput error}    & \multicolumn{2}{c}{Latency error}       & \multicolumn{2}{c}{Throughput error}    & \multicolumn{2}{c}{Latency error} \\
                            \cmidrule{2-9}
                & Downlink & \multicolumn{1}{c}{Uplink} & Downlink & \multicolumn{1}{c}{Uplink} & Downlink & \multicolumn{1}{c}{Uplink} & Downlink        & Uplink        \\
              \midrule
Max Packet size (\%) & 0.85  & \multicolumn{1}{c}{0.06}  & 0.07  & \multicolumn{1}{c}{0.27}  & 0.28  & \multicolumn{1}{c}{0.06}   & 1.48         & 0.23         \\
Mean Packet size (\%) & 0.59  & \multicolumn{1}{c}{0.13}  & 0.58  & \multicolumn{1}{c}{0.38}   & 0.35  & \multicolumn{1}{c}{0.13}  & 1.55         & 0.51         \\
              \bottomrule \\
        \multicolumn{9}{c}{Configuration B -- Multiple immersive UEs with PF}                                                                                           \\
              \cmidrule{2-9}
              & \multicolumn{4}{c}{Scenario A: Full offloading}                                    & \multicolumn{4}{c}{Scenario B: Deep learning offloading}                     \\
              \cmidrule{2-9}
              & \multicolumn{2}{c}{Throughput error}    & \multicolumn{2}{c}{Latency error}       & \multicolumn{2}{c}{Throughput error}    & \multicolumn{2}{c}{Latency error} \\
              \cmidrule{2-9}
              & Downlink & \multicolumn{1}{c}{Uplink} & Downlink & \multicolumn{1}{c}{Uplink} & Downlink & \multicolumn{1}{c}{Uplink} & Downlink        & Uplink        \\
              \midrule
Max Packet size (\%) & 0.57  & \multicolumn{1}{c}{0.23}  & 0.71  & \multicolumn{1}{c}{0.15}  & 1.55  & \multicolumn{1}{c}{0.16}  & 0.04         & 0.48         \\
Mean Packet size (\%) & 0.65  & \multicolumn{1}{c}{0.46}  & 0.29  & \multicolumn{1}{c}{0.32}  & 1.50  & \multicolumn{1}{c}{0.28}  & 0.03         & 0.36         \\
               \bottomrule \\
 \multicolumn{9}{c}{Configuration C -- Multiple immersive UEs with MT}                                                                                          \\
              \cmidrule{2-9}
              & \multicolumn{4}{c}{Scenario A: Full offloading}                                    & \multicolumn{4}{c}{Scenario B: Deep learning offloading}                     \\
              \cmidrule{2-9}
              & \multicolumn{2}{c}{Throughput error}    & \multicolumn{2}{c}{Latency error}       & \multicolumn{2}{c}{Throughput error}    & \multicolumn{2}{c}{Latency error} \\
              \cmidrule{2-9}
              & Downlink & \multicolumn{1}{c}{Uplink} & Downlink & \multicolumn{1}{c}{Uplink} & Downlink & \multicolumn{1}{c}{Uplink} & Downlink        & Uplink        \\
              \midrule
Max Packet size (\%) & 1.72  & \multicolumn{1}{c}{0.23}  & 0.83  & \multicolumn{1}{c}{0.34}  & 1.55  & \multicolumn{1}{c}{0.16}  & 0.06         & 0.20         \\
Mean Packet size (\%) & 1.28  & \multicolumn{1}{c}{0.46}  & 0.42  & \multicolumn{1}{c}{0.38}  & 1.92   & \multicolumn{1}{c}{0.28}  & 0.03         & 0.12  \\
\bottomrule
\end{tabular}
\end{table*}
All three configurations have in common the simulation parameters included in Table~\ref{tab:simparameters}. Each individual simulation run has a duration of 500 seconds and is repeated for each combination of configuration, number of UEs, offloading scenario (A and B), and type of data (synthetic with both packet types and captured data).  In all cases, there is a ``principal'' immersive UE closer to the emulated gNB than the other simulated UEs, from which we obtained the measurements used in this analysis. The goal is to study and compare the behavior of each type of IP traffic data at the application level, so we evaluated the application layer throughput and latency. The throughput is measured as the total mean throughput transmitted by all UEs. The latency, is measured only for the principal immersive UE. All the stochastic models, including the initial position of the non-principal UEs, have the same random seed across the experiments. The principal UE is placed 100~m away from the gNB to ensure it has priority regardless of the metric used for allocation, while the rest are placed randomly, at a longer distance.

The application layer mean throughput results obtained for the downlink transmission of Scenario~A for Configuration~C, are depicted in Fig.~\ref{fig:resultstp}. It is evident that the difference between the real, and the modeled data, for the total of UEs, is very low. We observe that from 8 UEs onward, the network starts saturating and the throughput does not increase linearly. This is because UEs with worse channel quality get fewer allocation grants. The measured latency behaves similarly showing low differences. Furthermore, similar results were obtained for all other configurations and scenarios. Overall, the throughput and latency differences between the captured and synthetic data obtained from FikoRE simulations are gathered in Table~\ref{tab:macroresults}. These differences are expressed by the relative mean error across emulation runs with different numbers of UEs. We observe that they are very low,  below $2\%$, in all cases. Besides, the differences between the Max Packet and Mean Packet cases of the IP packet sizes are negligible, with a mean difference of less than $0.04\%$. These results validate the goodness of the fitting of the proposed models for application-level simulations.

As a further step, we assess how well the synthetic traffic data generated with our model behave on the lower layers of the stack compared to the captured data. In particular, we study the resource allocation differences when using the captured or synthetic data as input for the simulator. Besides, we highlight the necessity of an accurate model which includes also the inter-packet intervals, contrary to the models proposed in~\cite{3gpp_17}. In this context, we generated synthetic data using a simple Normal model using the statistical metrics from the captured data included in Table~\ref{tab:mean_values}. However, instead of generating multiple IP packets within an RTP frame, we generated all the bits within the RTP frame in the same timestamp. By doing so we do not only highlight the necessity of an accurate model in terms of RTP frame size and inter-frame interval, but also the relevance of the inter-packet interval models. We refer to this simpler model as ``Norm'' model.

\begin{figure}[t]
\centering
\includegraphics[width=\linewidth]{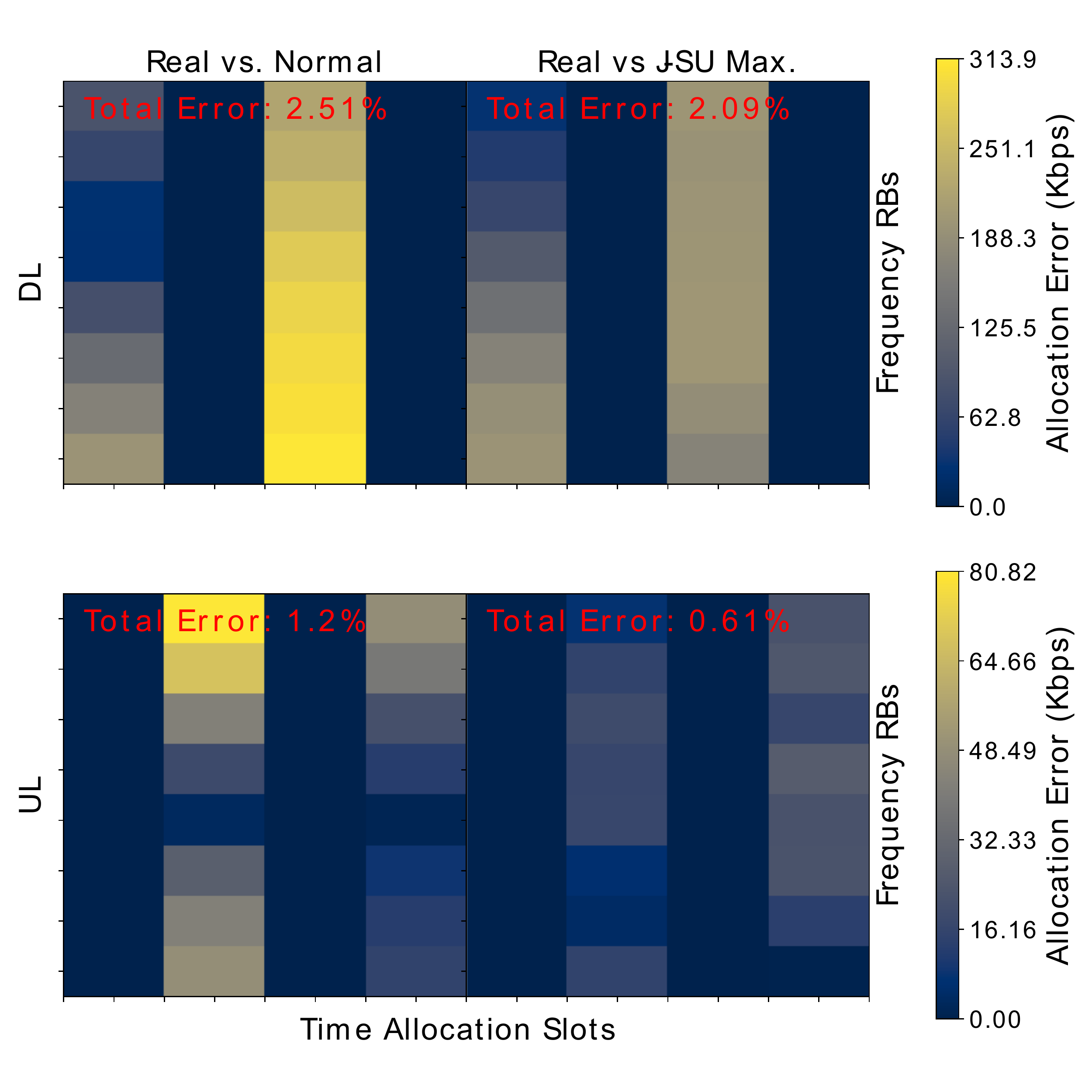}
\caption{Example of allocation error matrices for both transmission directions (UL and DL) between the captured and synthetic data. The error is estimated for the entire grid. This case corresponds to Scenario A and Configuration~B. }
\label{fig:allocationerrormat} 
\end{figure}

For this validation step, we also used FikoRE which is capable of logging every single allocation step, that is,  how the resources are allocated at each subframe. We use this information to compare the differences in terms of the allocated throughput, for each resource block (RB) within the allocation grid. More specifically, we measure the number of bits allocated for each RB and each UE. The number of RBs along the time and frequency axes depend on the bandwidth and selected numerology. The allocation error is estimated by comparing the bits allocated to each RB and UE when using the synthetic data from different models and using the actual XR traffic. We can build, for each UE, the allocation matrices illustrated in Fig.~\ref{fig:allocationerrormat}. These matrices express the resource allocation differences, or allocation errors, between a selected model and the actual XR traffic in bits per second, so the metric does not depend on the total duration of the simulation run. To estimate the allocation error of the entire grid as a percentage of the total transmitted error, we use the formula 
\begin{equation}
    e(\%) = 100\frac{\sum_{i=1}^{K} \left |t_{c}(i) - t_{m}(i)) \right |}{\sum_{i=1}^{K} t_{c}(i)},
\end{equation}
where $t_{m}(i)$ and $t_{c}(i)$ denote the allocated throughput of the model being evaluated, and the captured data, respectively, for the $i$th RB ($1\leq i\leq K$) along the total simulation time, with $K$ the total number of RBs. To really understand how the different sources of traffic data are being allocated, we decided to simulate a single UE, the principal one, in each run. By doing this, we avoid the effects of the selected configuration (such as the allocation metric, UEs channel quality, etc.) that directly affect the resource allocation procedure and can lead to inaccurate conclusions. 

Using the same configuration parameters described in Table~\ref{tab:simparameters}, we tested multiple combinations of total bandwidth and numerology $\mu$, which directly affect how the resource allocation grid is built, for a single immersive UE. In specific, we tested bandwidths of 40~MHz with $\mu=1$, 100~MHz with $\mu=2$, 200~MHz with $\mu=2$, and 200, 400, 800~MHz with $\mu=3$. Each simulation run had a duration of 500 seconds. The simulations were repeated for each configuration, scenario (A and B), and source of data (captured, Jonhson's $S_U$ with Max Packet size, and Norm). The synthetic data generated using the Mean Packet size presented no evident differences with the Max Packet size option. 

\begin{figure}[t]
\centering
\includegraphics[width=\linewidth]{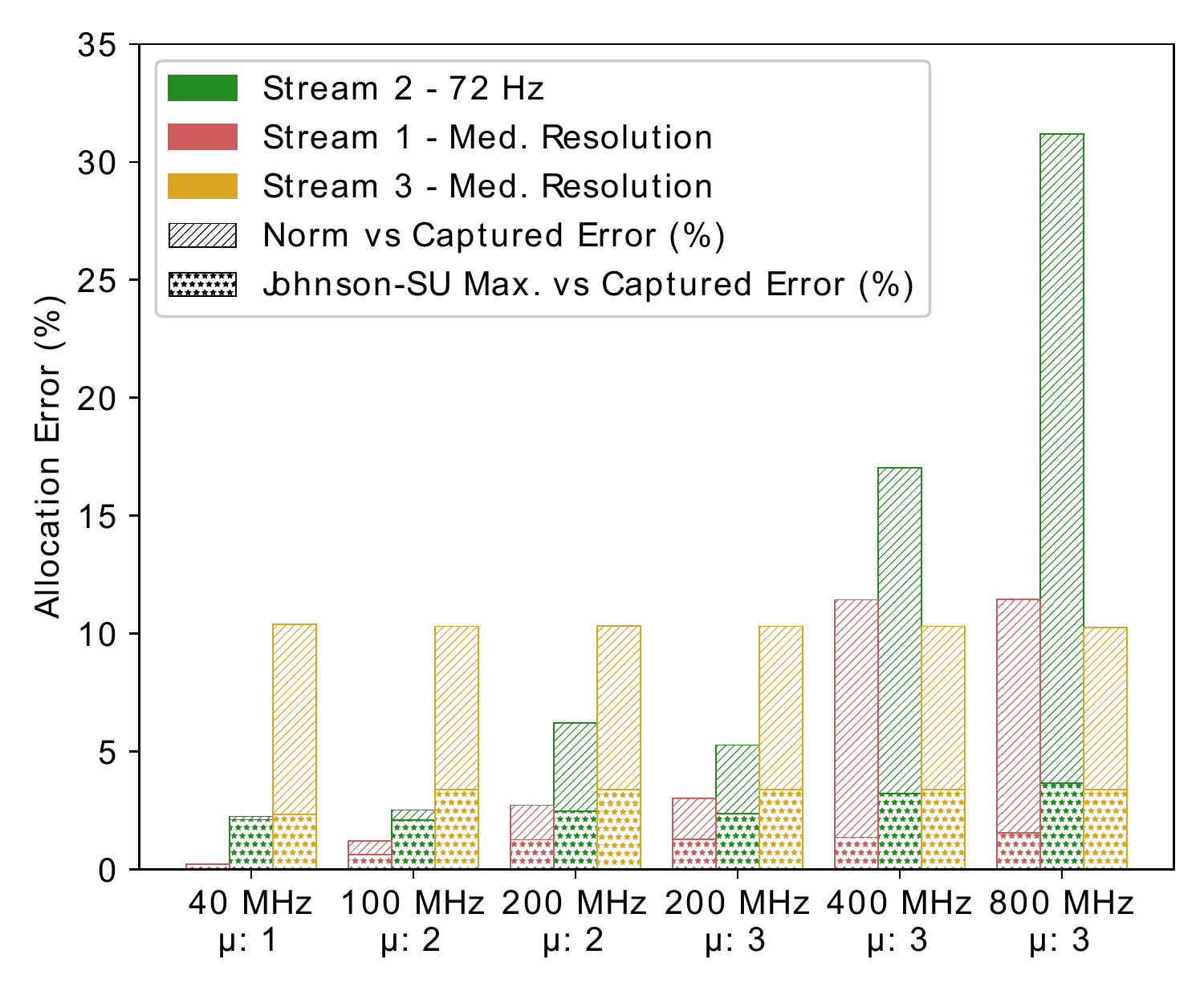}
\caption{Measured allocation errors between the captured and synthetic data using our emulation tool for each transmission direction (UL and DL) and validation scenario (A and B) for different numerology and bandwidth configurations. All simulations were done for a single UE. }
\label{fig:allocationerror} 
\end{figure}

Observing the measured allocation errors depicted in Fig.~\ref{fig:allocationerror} we can extract several conclusions. First, we notice that the allocation error is considerably higher for the Norm simpler model compared to the proposed Johnson's $S_U$ model. Besides, the error difference increases rapidly in favor of Johnson's $S_U$ model as we configure the emulator with more total bandwidth. Increasing the numerology also negatively impacts the performance of the Norm model. For low bandwidths, the error difference is low, as an RTP frame does not fit in a single subframe and has to be transmitted along several subframes. Therefore, the entire resource allocation grid gets saturated and the allocation differences, being estimated in comparison with the total allocated throughput in each RB, become hard to measure. On the contrary, for higher bandwidths, not all the RBs are allocated for each RTP frame and the differences become more noticeable. Our intuition is that the difference that we observe for higher bandwidths could also be observed if we could discard the saturated subframes. In addition, the allocation error of the proposed Johnson's $S_U$ models remains almost constant along the test configurations, which clearly is not the case for the Norm model. Thus, we get a strong hint of the importance of obtaining accurate models which include the inter-packet intervals, especially for high numerologies and bandwidths, in designing successful resource allocation techniques. 


\section{Conclusions}
\label{section:conclusion}
This work provided realistic traffic traces and associated models for XR offloading scenarios to complement and improve upon the models proposed in previous works, such as~\cite{3gpp_17}. We proposed two XR offloading scenarios that are at the cutting edge of the current state of the art. The first scenario represents a full offloading solution in which the XR HMD captures and transmits sensor data to a nearby server or MEC facility for processing and rendering ultra-high definition immersive frames, which are then transmitted back to the device. The second scenario focuses on offloading heavy ML algorithms, such as the real-time egocentric human body segmentation algorithm, which allows users to see themselves within the virtual scene.

The traffic data were captured using a recently introduced offloading architecture, described in~\cite{gonzalez2022arch}, with additional functionality presented in this work. To avoid any uncontrollable overhead that we did not aim to model, the data have been captured on the sender side, using a local host network. The IP traffic was captured for multiple resolutions for both the uplink and downlink streams and both offloading scenarios. 

The collected data were cleaned and post-processed, and we conducted a thorough analysis to determine the most appropriate modeling approach. We modeled the three main components of video traffic, that is, the frame size, inter-frame interval, and inter-packet interval, using continuous unimodal distributions. While many video or XR traffic models, such as~\cite{3gpp_17},  do not include inter-packet interval information, we consider it a crucial feature to include in XR traffic models, especially for resource allocation techniques design and optimization, as demonstrated in our validation experiments. We fitted multiple continuous distributions to the data for all resolutions and found that the Johnson's $S_U$ distribution provided the best fit, as determined by using the KS test.

The Johnson's $S_U$ distribution was fitted for all target parameters, scenarios, and resolutions. With these models, we generated synthetic data and used them in validation experiments with an open-source 5G RAN emulator~\cite{opensourcefikore}. These experiments compared the performance of the captured and synthetic data at both the application and resource allocation layers. At the application layer, we found that our models can generate realistic XR traffic data for the proposed scenarios. In the resource allocation layer, we demonstrated the importance of including inter-packet interval time for designing advanced resource allocation techniques specifically optimized for XR offloading.

In conclusion, the data and models presented in this work can be effectively used for the design, testing, improvement, and expansion of wireless network solutions in both academia and industry. They offer a comprehensive approach to studying extended reality (XR) offloading scenarios and provide insight into the importance of considering inter-packet interval times for resource allocation techniques. Overall, we believe that this work provides a useful contribution to the field of wireless networks and XR technology.
\appendix[XR IP Traffic Models Parameters]
\label{appendix}

The Johnson's $S_U$ distribution fitted parameters for the different types of streams and resolutions are summarized in Table~\ref{tab:final_models}. There fitted Johnson's $S_U$ distributions model the frame sizes, inter-frame intervals and inter-packet intervals of the captured RTP traffic. The given parameters can directly be used to generate realistic synthetic traffic. 

\setlength{\tabcolsep}{3pt}

\begin{table}[!h]
\centering
\caption{Fitted Johnson's $S_U$ distribution parameters for the RTP frame sizes, inter-frame intervals, and IP inter-packet intervals, along with the KS test values }
\label{tab:final_models}
\begin{adjustbox}{width=\linewidth}
\begin{tabular}{lrrrrr}
                                \multicolumn{6}{c}{Stream 1 -- Uplink Stereo Camera}  \\\midrule
                                    \multicolumn{6}{c}{Frame Size}  \\
                              \midrule
                               & \multicolumn{1}{c}{Location($\gamma$)} & \multicolumn{1}{c}{Scale($\delta$)} & \multicolumn{1}{c}{Shape-A($\lambda$)} & \multicolumn{1}{c}{Shape-B($\varepsilon$)} & \multicolumn{1}{c}{KS-Test} \\
                               \midrule

Low                            & 28205.00                               & 4990.40                               & -0.8691                                & 1.1236                                     & 0.0270                       \\
Med.                         & 73230.79                                & 12559.32                             & -0.7814                                 & 1.1801                                    & 0.0259                      \\
High                           & 220497.49                               & 20161.10                              & -0.4782                                & 1.2396                                     & 0.0192                      \\
\midrule

 \multicolumn{6}{c}{Inter-Frame Interval}                                                                                                                                                             \\\midrule

\multicolumn{1}{c}{}           & \multicolumn{1}{c}{Location($\gamma$)} & \multicolumn{1}{c}{Scale($\delta$)} & \multicolumn{1}{c}{Shape-A($\lambda$)} & \multicolumn{1}{c}{Shape-B($\varepsilon$)} & \multicolumn{1}{c}{KS-Test} \\\midrule

Low                            & 0.0168                                & 0.000204                            & 0.0282                                 & 1.2227                                    & 0.0025                      \\
Med.                         & 0.0168                                 & 0.000621                            & 0.1143                                 & 1.5517                                     & 0.0059                      \\
High                           & 0.0151                               & 0.007549                             & -0.6793                                 & 3.1580                                      & 0.0221                      \\\midrule

\multicolumn{6}{c}{Inter-Packet Interval}                                                                                                                                                            \\\midrule

\multicolumn{1}{c}{}           & \multicolumn{1}{c}{Location($\gamma$)} & \multicolumn{1}{c}{Scale($\delta$)} & \multicolumn{1}{c}{Shape-A($\lambda$)} & \multicolumn{1}{c}{Shape-B($\varepsilon$)} & \multicolumn{1}{c}{KS-Test} \\\midrule

Low                            & 1.61 E-6                                & 4.11 E-8                             & -1.2004                                & 0.4658                                    & 0.0820                       \\
Med.                         & 1.71 E-6                                & 4.08 E-8                             & -1.1691                                & 0.5173                                    & 0.0543                      \\
High                           & 1.86 E-6                                & 16.50 E-8                             & -1.4515                                 & 0.7052                                     & 0.0725                      \\\midrule\midrule
\\

     \multicolumn{6}{c}{Stream 2 -- Downlink Rendered Frames}  \\
     \midrule

\multicolumn{6}{c}{Frame Size}                                                                                                                                                                       \\\midrule

\multicolumn{1}{c}{}           & \multicolumn{1}{c}{Location($\gamma$)} & \multicolumn{1}{c}{Scale($\delta$)} & \multicolumn{1}{c}{Shape-A($\lambda$)} & \multicolumn{1}{c}{Shape-B($\varepsilon$)} & \multicolumn{1}{c}{KS-Test} \\\midrule

72~Hz & 216500.38                              & 52883.59                            & 0.1206                                  & 0.8444                                      & 0.0958                      \\
90~Hz & -136187.45                             & 14897.10                            & -9.7610                                  & 2.6910                                       & 0.0504                      \\\midrule

\multicolumn{6}{c}{Inter-Frame Interval}                                                                                                                                                             \\\midrule

\multicolumn{1}{c}{}           & \multicolumn{1}{c}{Location($\gamma$)} & \multicolumn{1}{c}{Scale($\delta$)} & \multicolumn{1}{c}{Shape-A($\lambda$)} & \multicolumn{1}{c}{Shape-B($\varepsilon$)} & \multicolumn{1}{c}{KS-Test} \\\midrule

72~Hz & 0.0139                                 & 4.37 E-5                            & -0.1091                                & 1.4920                                    & 0.0165                      \\
90~Hz & 0.0111                                  & 2.30 E-5                             & -0.1329                                 & 1.5173                                      & 0.0121                      \\\midrule

\multicolumn{6}{c}{Inter-Packet Interval}                                                                                                                                                            \\\midrule

\multicolumn{1}{c}{}           & \multicolumn{1}{c}{Location($\gamma$)} & \multicolumn{1}{c}{Scale($\delta$)} & \multicolumn{1}{c}{Shape-A($\lambda$)} & \multicolumn{1}{c}{Shape-B($\varepsilon$)} & \multicolumn{1}{c}{KS-Test} \\\midrule

72~Hz & 1.59 E-6                                & 5.40 E-8                             & -1.2095                                 & 0.5422                                     & 0.0548                      \\
90~Hz & 1.62 E-6                                & 5.31 E-8                             & -1.2583                                & 0.5210                                       & 0.0523                      \\\midrule\midrule
\\
     \multicolumn{6}{c}{Stream 3 -- Downlink Segmentation Masks}  \\
     \midrule
           \multicolumn{6}{c}{Frame Size}                                                                                                                                                                       \\
\midrule

\multicolumn{1}{c}{}           & \multicolumn{1}{c}{Location($\gamma$)} & \multicolumn{1}{c}{Scale($\delta$)} & \multicolumn{1}{c}{Shape-A($\lambda$)} & \multicolumn{1}{c}{Shape-B($\varepsilon$)} & \multicolumn{1}{c}{KS-Test} \\\midrule

Low                            & 4144.51                               & 1962.34                             & -0.4428                                & 1.4412                                     & 0.0177                      \\
Med.                         & 5547.72                               & 3879.20                            & -0.8572                                & 1.5868                                    & 0.0115                      \\
High                           & -3633.72                               & 16105.91                             & -3.8006                               & 2.9880                                     & 0.0220                       \\\midrule

\multicolumn{6}{c}{Inter-Frame Interval}                                                                                                                                                             \\\midrule

\multicolumn{1}{c}{}           & \multicolumn{1}{c}{Location($\gamma$)} & \multicolumn{1}{c}{Scale($\delta$)} & \multicolumn{1}{c}{Shape-A($\lambda$)} & \multicolumn{1}{c}{Shape-B($\varepsilon$)} & \multicolumn{1}{c}{KS-Test} \\\midrule

Low                            & 0.0168                               & 0.0001                             & -0.0092                               & 1.0634                                     & 0.0128                      \\
Med.                         & 0.0168                              & 0.0008                             & -0.0097                              & 1.6355                                   & 0.0106                      \\
High                           & 0.0164                                & 0.0023                             & -0.2288                                 & 1.1306                                     & 0.0227                      \\\midrule

\multicolumn{6}{c}{Inter-Packet Interval}           \\
\midrule

\multicolumn{1}{c}{}           & \multicolumn{1}{c}{Location($\gamma$)} & \multicolumn{1}{c}{Scale($\delta$)} & \multicolumn{1}{c}{Shape-A($\lambda$)} & \multicolumn{1}{c}{Shape-B($\varepsilon$)} & \multicolumn{1}{c}{KS-Test} \\\midrule

Low                            & 1.17 E-6                                & 1.91 E-8                             & -5.7054                                 & 0.9577                                     & 0.1805                      \\
Med.                         & 1.53 E-6                                & 6.07 E-8                             & -2.4255                                & 0.6070                                    & 0.1781                      \\
High                           & 2.14 E-6                                & 20.70 E-8                            & -0.9957                                & 0.4937                                     & 0.0788      \\
\bottomrule
\end{tabular}
\end{adjustbox}
\end{table}


%



\ifCLASSOPTIONcompsoc
  \section*{Acknowledgments}
\else
  \section*{Acknowledgment}
\fi

This work has received funding from the European Union (EU) Horizon 2020 research and innovation programme under the Marie Skłodowska-Curie ETN TeamUp5G, grant agreement No. 813391.

\ifCLASSOPTIONcaptionsoff
  \newpage
\fi



%
\bibliographystyle{ieeetr}
\bibliography{refs}


%

\iftrue                

\vskip -2\baselineskip plus -1fil 
\begin{IEEEbiographynophoto}{Diego~Gonz\'alez~Mor\'in} is a Ph.D student at Nokia Bell Labs Spain, enrolled with Universidad Carlos III de Madrid, Spain. He received his B.Sc.
and M.Sc. in industrial engineering from Universidad Politécnica de Madrid in 2015 and 2018 respectively. In 2018, he received his M.Sc. in systems, control and robotics from Kunliga Tekniska Hgskolan (KTH), Stockholm, Sweden. After receiving his M.Sc. degrees, he joined Ericsson Research’s Devices Technologies group as a researcher, where his research interest focused on augmented reality technologies, a field in which he holds three patents. In August 2019, he joined Nokia Bell Labs as a Ph.D. student. He is currently pursuing a Ph.D. focused on the application of ultra-dense networks for the implementation of distributed media rendering.
\end{IEEEbiographynophoto}
\vskip -2\baselineskip plus -1fil 
\begin{IEEEbiographynophoto}{Daniele~Medda} is a Ph.D student at the International Hellenic University of Thessaloniki, Greece. He received both his B.Sc. in Electrical and Electronic Engineering and his M.Sc. in Internet Technologies Engineering from the University of Cagliari (Italy) in 2018 and 2020, respectively. From 2018 to early 2021 he was a research assistant at the MCLab of the University of Cagliari, focusing on immersive data coding. He joined the International Hellenic University in 2021. His research interests are IEEE 802.11be networks, MAC layer optimization, ultra-dense networking and related standardization.
\end{IEEEbiographynophoto}
\vskip -2\baselineskip plus -1fil 
\begin{IEEEbiographynophoto}{Athanasios~Iossifides} is a Professor in the Department of Information and Electronic Engineering at the International Hellenic University (Greece). He received his diploma in Electrical Engineering and his Ph.D. from the Department of Electrical and Computer Engineering of the Aristotle University of Thessaloniki. From 1999 to 2010, he was with COSMOTE SA as a telecommunications engineer and the head of the Network Management Section of North Greece. He has served as Editor for Wiley Transactions on Emerging Telecommunication Technologies and IEEE Communications Letters and as a TPC co-chair or member in numerous international conferences. He has participated in several national and international research projects on wireless communications, the Internet of Things, and STE(A)M education which comprise the main fields of his research interests.
\end{IEEEbiographynophoto}
\vskip -2\baselineskip plus -1fil 
\begin{IEEEbiographynophoto}{Periklis~Chatzimisios} serves as a Professor in the Department of Information and Electronic Engineering at the International Hellenic University (Greece). Moreover, he has been awarded the title of Researcher Professor by the University of New Mexico (USA). He is also a Visiting Fellow in the Faculty of Science \& Technology, Bournemouth University (UK). Dr. Chatzimisios is/has been involved in several standardization and IEEE activities under the IEEE Communication Society (ComSoc). He is currently the Chair of the IEEE ComSoc Young Professionals Standing Committee, Chair of the Communications Chapter \& Professional Activities for the IEEE Greece Section, and an active member of the IEEE Future Networks Initiative. His research interests include performance evaluation and standardization of mobile/wireless communications, Internet of Things, 5G/6G, Industry 4.0, Smart Cities, and vehicle networking.
\end{IEEEbiographynophoto}
\vskip -2\baselineskip plus -1fil 
\begin{IEEEbiographynophoto}{Ana~Garc\'ia~Armada} (S’96–A’98–M’00–SM’08) is a professor at University Carlos III of Madrid, Spain. She has published approximately 150 refereed papers and she holds four patents. She serves on the Editorial Board of IEEE Transactions on Communications and the Open Journal of the IEEE Communications Society. 
She has received several awards from University Carlos III of Madrid, including an excellent young researcher award and an award for best practices in teaching. 
She was awarded the third place Bell Labs Prize 2014 for shaping the future of information and communications technology. She received the outstanding service award from IEEE ComSoc Signal Processing for Communications \& Computing Technical Committee (formerly SPCE) and the IEEE ComSoc/KICS Exemplary Global Service Award. Her research mainly focuses on signal processing applied to wireless communications.
\end{IEEEbiographynophoto}
\vskip -2\baselineskip plus -1fil 
\begin{IEEEbiographynophoto}{Alvaro~Villegas} leads the Extended Reality Lab in Nokia, a research center focused in the application of immersive media (VR, AR, XR) to human communications. He received a six-year telecommunications engineering degree at Universidad Polit\'ecnica de Madrid (Spain) and he completed an MBA Core Program at ESCP Europe Business School. Alvaro received the Distinguished Member of Technical Staff title from Bell Labs. He has dedicated his nearly 30 years of professional life to innovate in digital video in different companies: Telefonica, ONO, Motorola, Nagravision, Alcatel-Lucent and Nokia, where he has filed more than 40 patents. In his former role as Head of Bell Labs in Nokia Spain and now as lead of XR Lab he applies XR, AI/ML and 5G/6G technologies to improve human communications.
\end{IEEEbiographynophoto}
\vskip -2\baselineskip plus -1fil 
\begin{IEEEbiographynophoto}{Pablo~Per\'ez} is Lead Scientist at Nokia Extended Reality Lab (Madrid, Spain). He is Telecommunication Engineer (BSc+MSs, 2004) and PhD in Telecommunications (2013) from Universidad Polit\'ecnica de Madrid, Spain, and Nokia Distinguished Member of Technical Staff (2022). He has worked as R\&D engineer of digital video products and services in Telefonica, Alcatel-Lucent and Nokia; as well as a researcher in future video technologies in Nokia Bell Labs. He is currently leading the scientific activities of Nokia XR Lab, addressing the end-to-end technological chain of the use of Extended Reality for human communication: networking, system architecture, processing algorithms, quality of experience and human-computer interaction.
\end{IEEEbiographynophoto}
\fi





\end{document}